\documentclass[%
 aip,
 amsmath,amssymb,
 reprint,%
]{revtex4-1}

\usepackage{graphicx}% Include figure files
\usepackage{dcolumn}% Align table columns on decimal point
\usepackage{bm}% bold math
\usepackage[utf8]{inputenc}
\usepackage[T1]{fontenc}
\usepackage{mathptmx}
\usepackage{comment}
\usepackage{subfig}

\begin{document}

\preprint{AIP/123-QED}

\title[]{Performance and uniformity of a kilo-pixel array of Ti/Au transition-edge sensor microcalorimeters}
% Force line breaks with \\

\author{E. Taralli}
\email[Author to whom correspondence should be addressed: ]{e.taralli@sron.nl.}
\author{M. D'Andrea}
\author{L. Gottardi}
\author{K. Nagayoshi}
\author{M. L. Ridder}
\author{M. de Wit}
\author{D. Vaccaro}
\author{H. Akamatsu}
\author{M. P. Bruijn}
\affiliation{SRON Netherlands Institute for Space Research, Sorbonnelaan 2, 3584 CA Utrecht, The Netherlands.}%Lines break automatically or can be forced with \\

\author{J. R. Gao}
\affiliation{SRON Netherlands Institute for Space Research, Sorbonnelaan 2, 3584 CA Utrecht, The Netherlands.}%Lines break automatically or can be forced with \\
\affiliation{Faculty of Applied Science, Delft University of Technology, 2600 AA Delft, The Netherlands.}%Lines break automatically or can be forced with \\

\date{\today}% It is always \today, today,

\begin{abstract}
Uniform large transition-edge sensor (TES) arrays are fundamental for the next generation of X-ray space observatories. These arrays are required to achieve an energy resolution $\Delta E$ < 3 eV full-width-half-maximum (FWHM) in the soft X-ray energy range. 
We are currently developing X-ray microcalorimeter arrays for use in future laboratory and space-based X-ray astrophysics experiments and ground-based spectrometers.\\
In this contribution we report on the development and the characterization of a uniform 32$\times$32 pixel array with 140$\times$30 $\mu$m$^2$ Ti/Au TESs with Au X-ray absorber. 
We report upon extensive measurements on 60 pixels in order to show the uniformity of our large TES array. The averaged critical temperature is $T_\mathrm{c}$ = 89.5$\pm$0.5 mK and the variation across the array ($\sim$1 cm) is less than 1.5 mK. We found a large region of detector's bias points between 20\% and 40\% of the normal-state resistance where the energy resolution is constantly lower than 3 eV. In particular, results show a summed X-ray spectral resolution $\Delta E_\mathrm{FWHM}$ = 2.50$\pm$0.04 eV at a photon energy of 5.9 keV, measured in a single-pixel mode using a frequency domain multiplexing (FDM) readout system developed at SRON/VTT at bias frequencies ranging from 1 to 5 MHz.  
Moreover we compare the logarithmic resistance sensitivity with respect to temperature and current ($\alpha$ and $\beta$ respectively) and their correlation with the detector's noise parameter $M$, showing an homogeneous behaviour for all the measured pixels in the array.
 
\end{abstract}

\maketitle
%%%
%\begin{quotation}
%The ``lead paragraph'' is encapsulated with the \LaTeX\ 
%\verb+quotation+ environment and is formatted as a single paragraph before the first section heading. 
%(The \verb+quotation+ environment reverts to its usual meaning after the first sectioning command.) 
%Note that numbered references are allowed in the lead paragraph.
%
%The lead paragraph will only be found in an article being prepared for the journal \textit{Chaos}.
%\end{quotation}

\section{\label{sec:intro}Introduction}

Large and uniform array of detectors, with high spectral resolution (SR = $\Delta E/E$), are highly demanded for a number of scientific objectives in space observation and are getting crucial for the next generation of space observatories with large telescopes. Superconducting microcalorimeter devices like transition-edge sensors (TESs) \cite{IrwinHilton} are capable to deliver an SR > 2000 becoming the leading choice in most of the instruments.\\
TES sensors are very sensitive thermometers which are able to detect radiation in a wide energy range, e.g. from $\gamma$-ray down to submillimeter\cite{Ullom2,Ullom,minussi,brink,lolli,cabrera,fukuda,obrient,li}. TES consists of a single layer of superconducting material or of a multilayer of materials where a metallic intermedium is included to tune the critical temperature of the entire detector. A TES is weakly thermal coupled to the thermal bath ($T_\mathrm{bath} < T_\mathrm{c}$). This weak thermal link is often created via suspension of the TES on a SiN$_x$ membrane. Depending on the energy of the photons involved, an absorber, optical cavity or antireflection coating is used to achieve the required quantum efficiency, allowing the measurement of the energy or power of the incoming photons. The TES is self-heated to within its very steep superconducting-to-normal phase-transition by Joule power supplied by a voltage-bias circuit. An electro-thermal feedback insures a self-regulation in a selected working point\cite{IrwinHilton}. When a photon or a particle is absorbed, it generates an increase of the temperature causing a subsequent change in the resistance and therefore in the current flowing through the TES. This signal is read out using inductively coupled superconducting quantum interference devices (SQUIDs).\\
In the last decades, large arrays of TES microcalorimeters have been used for spectral imaging acquisition and are under development in a number of space telescopes\cite{spica,athena,hubs}. Specifically, SRON TES array has been selected as the backup option for the X-ray Integral Field Unit (X-IFU)\cite{xifu}which is one of the two instruments on board of the Athena space mission\cite{athena} and consists of an array of over 3000 TESs. It follows that the uniformity of such large arrays plays a crucial role in most of the applications. Any non-uniformities across the array could lead to different detectors having different energy resolution and/or response time, which would be inconvenient for an imaging-spectroscopy camera.\\
Fabrication of large arrays passes through a number of steps and is done in multiple layers as well as involves a wide deposition area. Because of this, each pixel may not be impacted equally, leading potentially to degrade the achievable uniformity of performance across an array. The capability to bias all the pixels approximately at the same working point, aiming to get the same performances over the full array, is of course very important for the quality and the reproducibility of an experiment. For these reasons, all the parameters that tend to affect the performance of the single detector and thus the homogeneity of the whole array have to be investigated. Those include for instance critical temperature, linearity of the transition curve, thermal conductance to the bath and quality of the interface between leads and bilayer.\\
The readout technology used in this work is frequency domain multiplexing (FDM)\cite{fdm1,fdm2}. It applies a set of sinusoidal AC carriers, which bias the TES detectors at their working points and are amplitude modulated when the TES detectors are hit by photons. The detectors are separated in frequency by placing them in series with $LC$ resonators, each having a specific resonant frequency. The frequency bands assigned to the detectors are separated to prevent the detectors from interacting with each other. This allows the readout of multiple TES pixels by one amplifier channel, which uses only one set of SQUID current amplifiers. We are currently using an 18-channel FDM readout system with bias frequencies between 1-5 MHz, which is a prototype version of the anticipated 40-channel FDM readout.\\
The paper is organized as follows: Sec.~\ref{sec:device} gives a quick overview of the modifications implemented in the last years in the pixel design, inside the TES community in general and in particular at SRON, to further improve TES performances; Sec.~\ref{sec:exp} describes the characteristics of the pixel introduced in the kilo-pixel array (Sec.~\ref{sec:fab}) and the experimental setup used during this work (Sec.~\ref{sec:setup}); Sec.~\ref{sec:uniformity} presents the measurements performed on the various pixels on the TES array, in particular the uniformity tests such as critical temperature and thermal conductance from one side of the chip to the other (Sec.~\ref{sec:tc}), the partial logarithmic derivatives of TES resistance $R(T, I)$ with respect to temperature and current, $\alpha$ and $\beta$, respectively and the noise factor  $M$ (Sec.~\ref{sec:impedance}), noise equivalent power (Sec.~\ref{sec:nep}) and single pixel energy resolution (Sec.~\ref{sec:deltae}); In  Sec.~\ref{sec:conclusion}, we draw  the main conclusion about the uniformity of such kilo-pixel array.

\section{\label{sec:device}Pixel evolution under AC-bias}
\begin{figure*}
%\subfloat[]{\includegraphics[width=0.445\linewidth, keepaspectratio]{iv_phase_vs_r0_LCU087.png}\label{fig:IVcurves}}% Here is how to import EPS art
\includegraphics[width=0.445\linewidth, keepaspectratio]{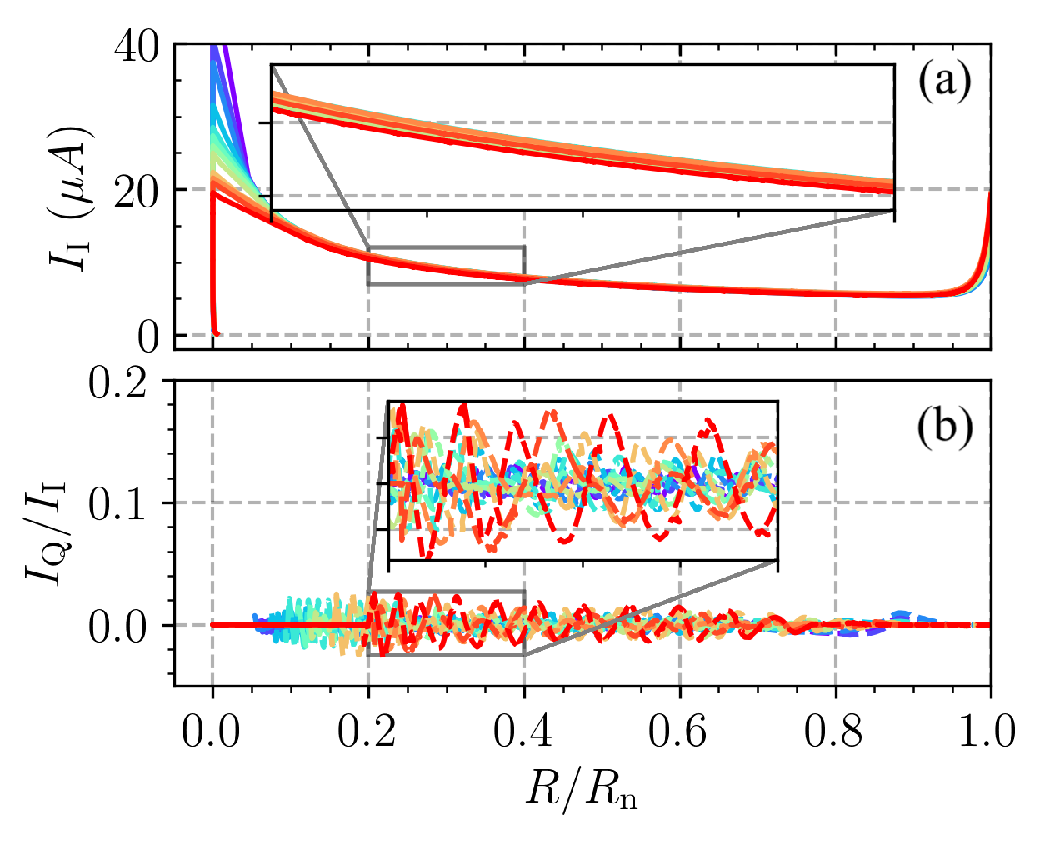}% Here is how to import EPS art
\includegraphics[width=0.58\linewidth, keepaspectratio]{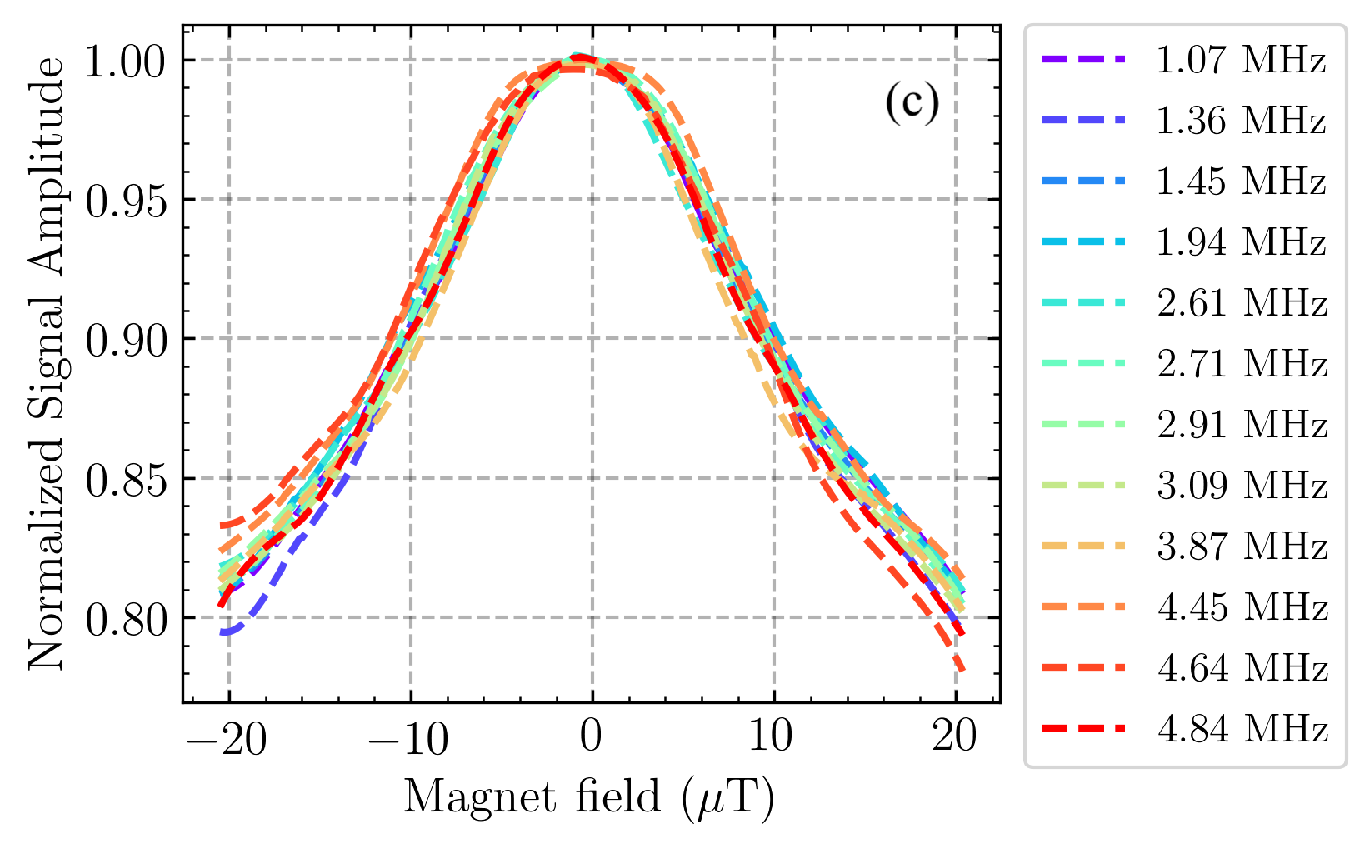}% Here is how to import EPS art
%\subfloat[]{\includegraphics[width=0.58\linewidth, keepaspectratio]{Bscan_vs_mag_withlegend.png}\label{fig:Bscan}}% Here is how to import EPS art
%\subfloat[]{\includegraphics[width=0.1\linewidth, keepaspectratio]{legend.png}}% Here is how to import EPS art
\caption{\label{fig:IV_B}TES current in-phase (a) and the ratio between the quadrature and in-phase current $I_\mathrm{Q}/I_\mathrm{I}$ (b), as a function of $R/R_\mathrm{n}$ for 12 pixels.  (c) Detector's current (dashed lines) as a function of the perpendicular magnetic field applied on the same 12 pixels when biased in the same bias point $R/R_\mathrm{n} \sim$ 30\%.}
\end{figure*}

In the last decade, TES pixel design has passed through a number of changes. The matching between the readout techniques \cite{fdm1,fdm2,mwave,cdm,tdm} and the detector mainly drives the pixels design in order to get the best performance from the whole array. For instance, the connection of the bilayer to the higher $T_\mathrm{c}$ superconducting Nb leads \cite{sadleir,ridder} and the presence of normal metal structures in the bilayer, cause proximity effects that strongly affect the TES behaviour both under DC \cite{smith} and AC bias \cite{taralli}. The TES behaves typically as an SS'S or SNS junction composed by two superconductors S and S' (where $T_\mathrm{c'}<T_\mathrm{c}$) or one superconductor S and one normal conductor N, respectively. The Josephson effects are regularly observed both in the response to the perpendicular magnetic field and, under AC bias, in the changes of the TES reactance across the superconducting transition. It leads to a considerable variation (Fraunhofer-like oscillations) of the device's critical current as a function of the magnetic field $B$. At the same time, the transition curve is characterised by kinks, the location of which depends, among other things, on the alignment of the normal metal structures on the bilayer\cite{nick2}. Uniform and predictable tuning of TESs across a large array is simplified by the minimization of these kinks and oscillations in the TES transitions.\\ 
Added to this, biasing TES microcalorimeters in AC introduces another issue that has to be taken into account during the pixel design iterations. In AC-biased low-ohmic TES microcalorimeters, the Josephson effects are masked by another frequency-dependent dissipation mechanism, on which we have already reported before \cite{gottardi1,gottardi2} and which is related to the generation of eddy currents in the normal metal structures surrounding the TES.\\
The measurement of the quadrature component of the $IV$ curves in AC-biased TESs and the dependency of the detector's current as a function of the magnetic field, are useful methods to quantify most of these effects. The quadrature component of $IV$ curves shows an oscillatory behaviour dependent on the driving bias frequency and generally, the period and the amplitude of the oscillations decreases with the bias frequency. Moreover, the amplitude is larger at low bias voltages due to the fact that the Josephson current is reduced when  $\sqrt{PR_\mathrm{n}}/F_\mathrm{b}$ increases, where $P$ is the detector power, $R_\mathrm{n}$ is TES normal resistance and $F_\mathrm{b}$ is the bias frequency\cite{gottardi2,gottardi3}. High saturation power, high normal resistance and low bias frequency readout is one way to minimise the weak-links in the TES. The saturation power is related to the $G$ of the detector, which is typically constrained by the scientific application while the bias frequency by the engineering requirements. The only parameter left free for the optimisation is the TES normal resistance. At the same time, the thermal fluctuation noise internal to the microcalorimeter \cite{nick} increases as the thickness of the superconducting bilayer decreases. This affects how large the $R_\mathrm{n}$ can be designed for optimal performance for a given application.\\ Recently a new generation of TES designs \cite{ken} made with a thicker bilayer but with high aspect-ratio (to keep a larger TES's normal resistance) have shown a smaller Josephson current under AC bias giving high energy resolution performance \cite{taralli2,martin}. 
We report some specific data in this section in order to show the effect of these changes on the TES pixel design with an aspect-ratio (length$\times$wide) 140$\times$30 $\mu$m$^2$ placed in the large array subject of this work. Josephson effects can be quantified by measuring the $IV$ curves for 12 pixels connected at different bias frequencies and in particular by looking at the ratio $I_\mathrm{Q}/I_\mathrm{I}$ between the detector quadrature and in-phase current, as shown in Fig.~\ref{fig:IV_B}a and Fig.~\ref{fig:IV_B}b. Besides the uniformity of the detectors current as a function of the bias voltage, it is worth noting how $I_\mathrm{Q}/I_\mathrm{I}$  ratio remains less than 2\% low in transition even at high bias frequency. This is  an improvement if compared with the previous SRON devices\cite{gottardi1}:  $I_\mathrm{Q}/I_\mathrm{I}$<25\% in low-ohmic and low-power detector and  $I_\mathrm{Q}/I_\mathrm{I}$<5\% in high-power and high normal resistance detectors. Fig.~\ref{fig:IV_B}c shows the variation of the device current as a function of the magnetic field at $T_\mathrm{bath}$ = 55 mK where all the TESs measured have been biased at the same working point such that $R/R_\mathrm{n} \sim$ 30\%. For this specific TES geometry, we expect that the Fraunhofer-like oscillations of the device current has a period of $\sim$0.5 $\mu$T. From Fig.~\ref{fig:IV_B}c we don't see any modulation of the TES current versus the applied magnetic field up to $\sim$8 $\mu$T. It means that this TES is very weakly affected by the proximity effect. Other details about the uniformity and the impact of the weak-links on this large array are being reported in the following of this paper. Iterations to further improve the pixel design will be continuing and are still needed. Increasing the thickness of the bilayer to improve the device thermalisation, enhance more the aspect ratio of the detector to keep an high normal resistance and finding new materials for the leads are only fews of the advances that will be faced in the coming future. 

\section{\label{sec:exp}Array description and experimental details}

\subsection{\label{sec:fab}Kilo-pixel array}
\begin{figure}
\includegraphics[width=1\linewidth, keepaspectratio]{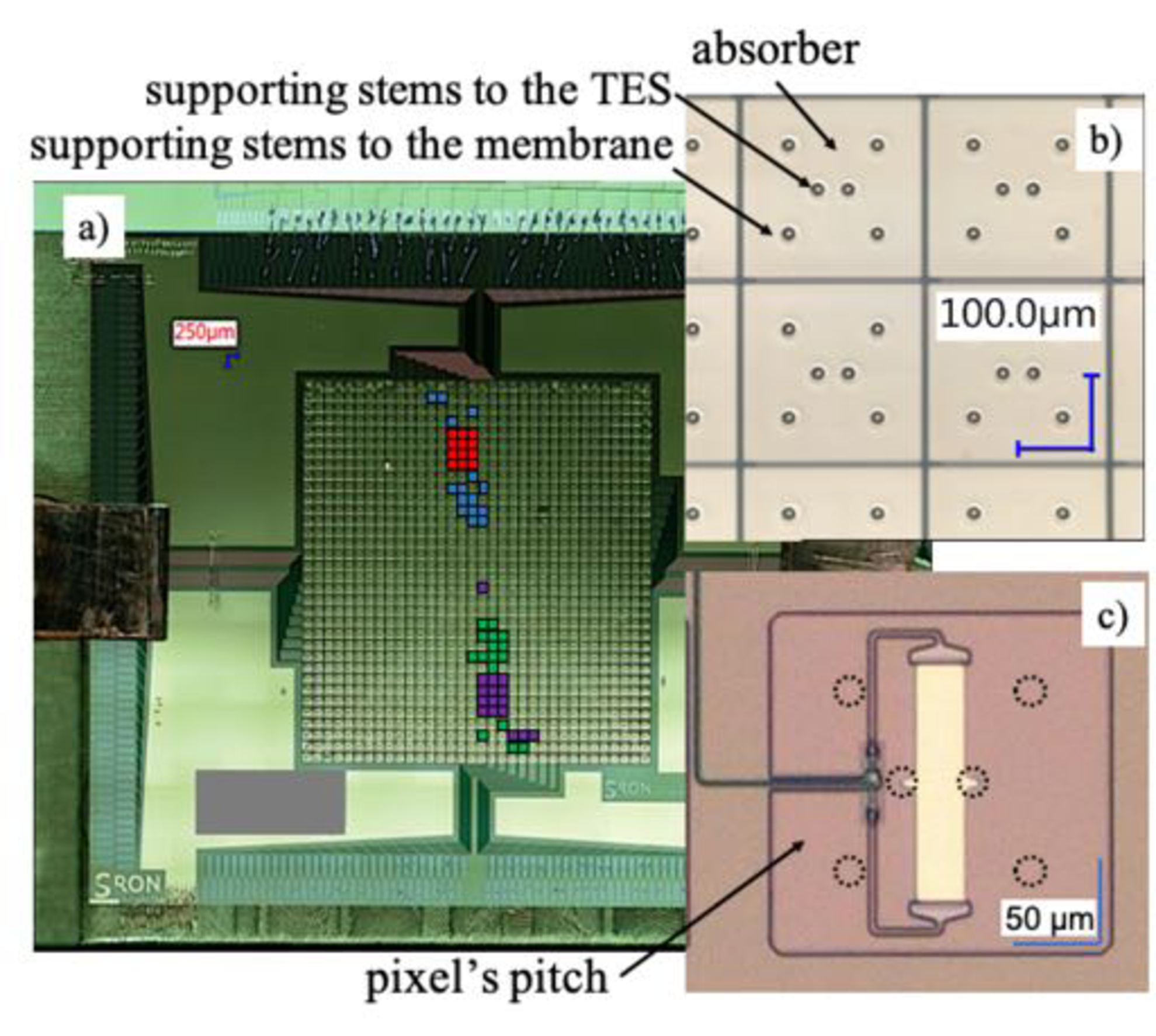}\\% Here is how to import EPS art
\caption{\label{fig:array} (a) Top view of 32$\times$32 kilo-pixels array with the pixels measured during the 4 Runs: purple (Run1), green (Run2), blue (Run3) and red (Run4). (b) Picture of the X-ray absorbers located in the 32$\times$32 array. The absorber size is 240$\times$240 $\mu$m$^2$ with a gap of 10 $\mu$m between the neighbors and six supporting stems. (c) Picture of a 140$\times$30 $\mu$m$^2$ Ti/Au TES (taken before absorber deposition) connected to a microstrip line via interconnecting leads. Dotted circles identify the area where the supporting stems will eventually be grown. The area indicated by the arrow is the membrane area (roughly, it is where the SiO$_\mathrm{2}$ layer is removed).}
\end{figure}

The kilo pixel array under test has been fabricated on an 4 inch wafer together with a number of 5$\times$5 uniform and mixed arrays and with other 32$\times$32 uniform kilo pixel arrays with various aspect-ratio (length$\times$width) devices. Fig.~\ref{fig:array}a shows the 32$\times$32 Ti/Au TES array and in particular the 60 pixels that have been measured during four different measurement cycles (from now on called Runs). All the TESs use a 0.5 $\mu$m thick low-stress, silicon-rich silicon nitride SiN$_x$ membrane and have the same bilayer thickness of Ti (35 nm) and Au (200 nm) resulting in a normal resistance per square $R$ = 26.2 m$\Omega$/$\square$ and critical temperature $T_\mathrm{c} \sim$  90 mK. All the absorbers consist of  Au (2.3 $\mu$m thick) and have the same size (240$\times$240 $\mu$m$^2$)  with a heat capacity $C$ = 0.85 pJ/K at $T_\mathrm{c}$. Each absorber has four supporting stems directly connected to the membrane and two stems directly connected to the bilayer as shown in Fig.~\ref{fig:array}b. The design of the  device with dimension (length$\times$width) 140$\times$30 $\mu$m$^2$ that has been explored in this work is shown in Fig.~\ref{fig:array}c with a normal resistance $R_\mathrm{n}$ = 121 m$\Omega$ and an expected thermal conductance \cite{thermalG1} $G\sim$ 95 pW/K at $T_\mathrm{c}$. More details on fabrication of such SRON TES arrays have been published in a previous work \cite{ken}.

\subsection{\label{sec:setup}Experimental set-up}
\begin{figure}
\includegraphics[width=1\linewidth, keepaspectratio]{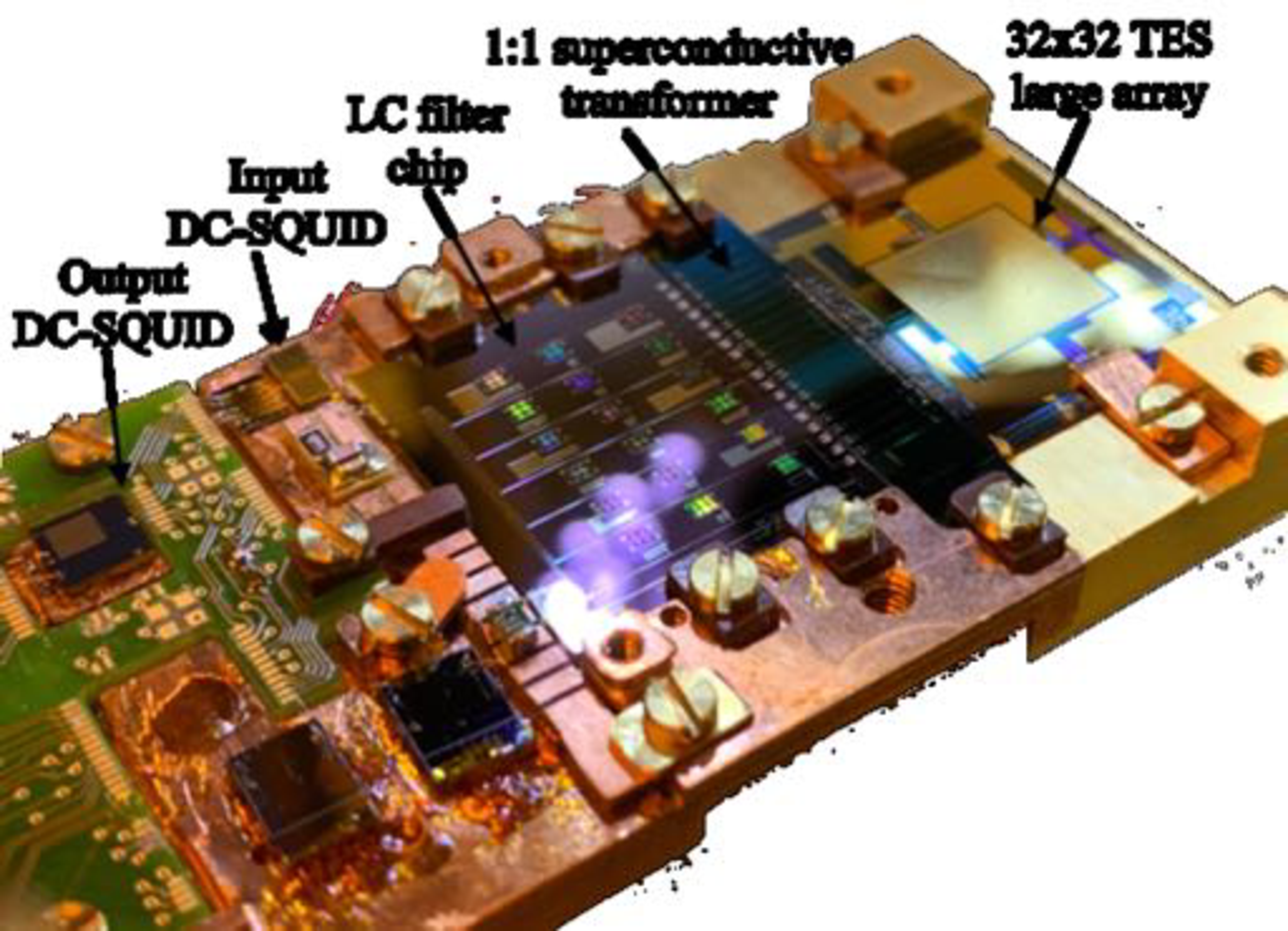}\\% Here is how to import EPS art
\caption{\label{fig:setup} Picture of the setup used to characterise the kilo-pixels array, hanged at the mixing chamber of a dilution refrigerator and held at a base temperature of 50 mK. Main parts are highlighted by arrows.}
\end{figure}

The characterization of the kilo-pixel arrays was performed in an experimental measurement set-up named XFDMLarge and it is shown in Fig.~\ref{fig:setup}. It was installed in a dilution refrigerator that can provide a bath temperature of $\sim$ 40 mK. TESs were characterised under AC bias using an existing FDM readout system (1-5 MHz) \cite{fdm1} in single-pixel mode configuration, where only one device is biased at a time, and all others are left in the superconducting state. Each TES is connected in series with an $LC$ resonator on an $LC$ filter chip with a coil inductance $L$ = 2 $\mu$H and a 1:1 transformer chip. The kilo-pixel TES array chip and the cryogenic components of FDM readout were mounted in a low magnetic impurity copper bracket fitted into an Al superconducting shield. The bracket also accommodates a heater, a thermometer and a Helmholtz coil. The heater and thermometer are used to stabilise the temperature locally on the chip. The coil is for applying a uniform magnetic field perpendicular to the TES array to compensate any remnant magnetic field trapped in the experiment set-up.

\section{\label{sec:uniformity}Uniformity characterization}

\subsection{\label{sec:tc}Critical temperature and thermal conductance}
\begin{figure}
\includegraphics[width=1\linewidth, keepaspectratio]{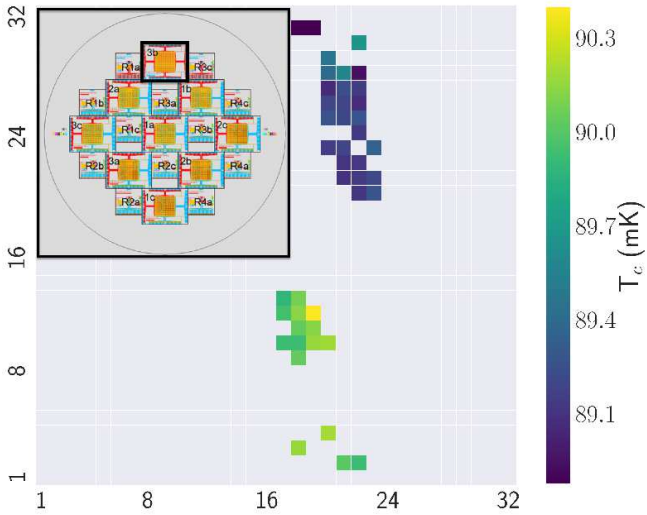}\\% Here is how to import EPS art
\caption{\label{fig:tc} Heat map of the critical temperature measured all over the kilo-pixel array. The inset shows the location of the array in the whole wafer indicated in the black box. Run1 has been neglected due to a different warm electronics used in this measurement.}
\end{figure}

\begin{figure}
\includegraphics[width=1\linewidth, keepaspectratio]{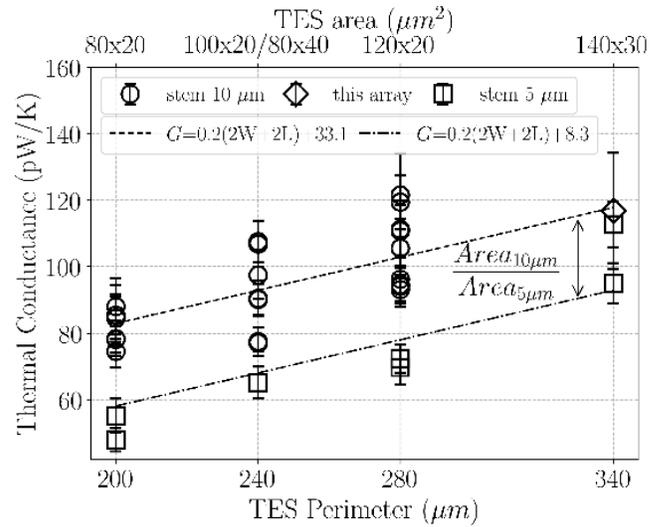}\\% Here is how to import EPS art
\caption{\label{fig:g} Thermal conductance of TESs with different aspect-ratio (perimeters) and 10 $\mu$m stem's diameter to support the absorber: 140$\times$30 (340 $\mu$m) placed in this array and averaged over 44 pixels (diamond point), 80$\times$20 (200 $\mu$m), 80$\times$40 and 100$\times$20 (200 $\mu$m) and 120$\times$20 (280 $\mu$m) measured in another setup (open dots). Fit (dashed line) follows the dependency of the thermal conductance from perimeter ($2W+2L$). Same aspect-ratios have been measured on another mixed array with 5 $\mu$m stem's diameter to support absorbers (open squares). Fit shifted down of ratio between the areas of the two different stems (dash-dot line). }
\end{figure}

Any variation in the transition temperature of pixels across the array affects the uniformity of the energy resolution, the bias point and the speed of the detectors over the whole array. Achieving a sufficiently homogeneity both in the thickness of the bilayer and in the subsequent processing stages over the full array, guarantees to bias all devices approximately in the same sensitive part of the transition aiming to the same detector's performances. \\
By measuring all the $IV$ curves for the selected set of pixels at different bath temperatures $T_\mathrm{bath}$, we are able to calculate the TES dissipated power $P_\mathrm{TES}$, for example at the minimum of the $IV$ curve, as a function of $T_\mathrm{bath}$. We fit these data using the balancing between the dissipated TES electrical power and the dissipated TES thermal power to the bath $P_\mathrm{TES} = I_\mathrm{TES}^2\times R_\mathrm{TES}=K\times (T_\mathrm{c}^n-T_\mathrm{b}^n)$,  where $n$ is a number whose value depends on the dominant thermal impedance between the substrate and the electrons in the superconducting film and $K$ is a material and geometry dependent parameter. In this way we  determine the critical temperature $T_\mathrm{c}$ and the thermal conductance $G=\mathrm{d}P_\mathrm{TES}/\mathrm{d}T$ for each pixel.\\
In Fig. ~\ref{fig:tc} we highlight the uniformity of the array in terms of critical temperature by means of a heat map. We can easily notice how all the pixels belonging to the same quadrant, show a difference in the transition temperature less than 0.6 mK, while the total variation between the lower and upper quadrant over the whole array  ($\sim$ 1 cm) is less than 1.5 mK. We would like to mention that the first 16 pixels in Run1 (purple squares in Fig. ~\ref{fig:array}a), are not included in this figure. In this run, a different warm electronics unit was used during the characterisation\cite{footnote}. We would like to underline that between two runs inside the same quadrant, only the wire-bondings changed, whereas between two runs in different quadrants the chip array rotated by 180 degrees as well. The position of the thermometer and of the other components of the setup has not been changed among the runs. Moreover, the same TES has been measured during Run3 and Run4 and we have obtained $T_\mathrm{c}$ = 89.6 mK and $T_\mathrm{c}$ = 89.4 mK, respectively. This temperature variance of 0.2 mK is smaller than 0.6 mK and 1.5 mK that we obtained inside one quadrant and between the two quadrants, respectively. For this reason we rule out the possibility that our characterization is mainly guided by calibration instability or non-repeatability of the measurement. Averaging all the critical temperatures for the 44 pixels we get a $T_\mathrm{c}$ = 89.5 with a standard deviation of 0.5 mK. The nature of this dispersion around the mean value is still under discussion. However,  it might be explained considering some of the critical aspects in the whole fabrication process, e.g.  the TES patterning by means of wet etching process, the stress on the membrane due to the absorber and the uniformity of the absorber itself.\\
The averaged thermal conductance that we have measured across the array is $G$ = 117$\pm$17 pW/K (diamond point in Fig. ~\ref{fig:g}). The thermal transport in the nitride membranes is quasi-ballistic because of extremely long phonon mean-free paths, as we have already shown in the past \cite{thermalG1}, resulting in a thermal conductance that depends on the perimeter of the TES and the thickness of the membrane. Comparing the value of our measured thermal conductance with earlier tests of similar devices fabricated by NASA Goddard, we get that our current value is about 20 pW/K above the expected value \cite{thermalG2, smith}. We found that the combined TES film plus supporting stems area determines the thermal conductance. In Fig.~\ref{fig:g} we report on the thermal conductance measured on TES with different geometry (80$\times$20, 100$\times$20, 80$\times$40, 120$\times$20 and 140$\times$30 $\mu$m$^2$) placed in two arrays with different area of the supporting stem (10 and 5 $\mu$m of diameter). These results scale consistently with the detector perimeter ($2W+2L$), but also show a clear shifting between them. The dashed line in Fig.~\ref{fig:g} is the linear fit of the open dots and the diamond point where $a(2W+2L)$ is the TES thermal conductance scaling with the  perimeter and $b$ is the thermal conductance due to the stems. Using the equation $G=n\Sigma \Omega T^{n-1}$ where $n$ is the exponent of power flow to the heat bath, $\Omega$ is the material volume and $\Sigma$ is a material-dependent constant\cite{IrwinHilton,thermalG1}, we can estimate the relative impact of the stems to the measured thermal conductance. Assuming that the perimeter of the stems is the only difference between the two arrays  (material, $n$ and height of the stems are the same), the thermal conductance of the 5 $\mu$m stems is nothing but the thermal conductance $b$ of the 10 $\mu$m stems, scaled down by the ratio between the stem's area, as shown by the dot-dashed line in Fig.~\ref{fig:g}. A further reducing of the stems diameter as well as the number of the stems directly connected to the membrane will reduce this additional thermal conductance.

\subsection{\label{sec:impedance}$\alpha$, $\beta$ and $M$ factor}

\begin{figure}
%\subfloat[]{\includegraphics[width=1.0\linewidth, keepaspectratio]{alpha_LCUrun84_few}\label{fig:alpha}}\\% Here is how to import EPS art
\includegraphics[width=1.0\linewidth, keepaspectratio]{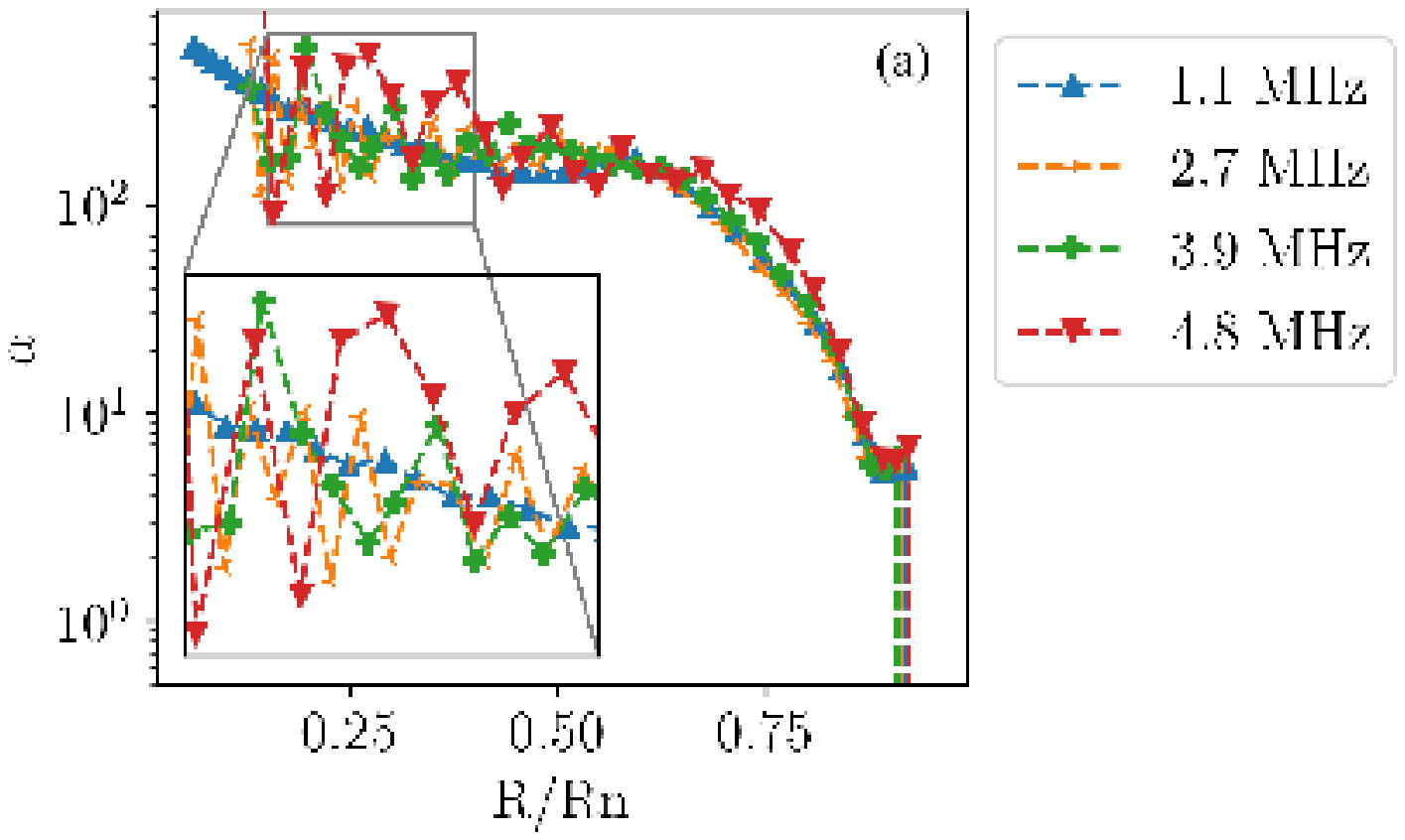}\\% Here is how to import EPS art
%\subfloat[]{\includegraphics[width=1.0\linewidth, keepaspectratio]{beta_LCUrun84_few}\label{fig:beta}}\\% Here is how to import EPS art
\includegraphics[width=1.0\linewidth, keepaspectratio]{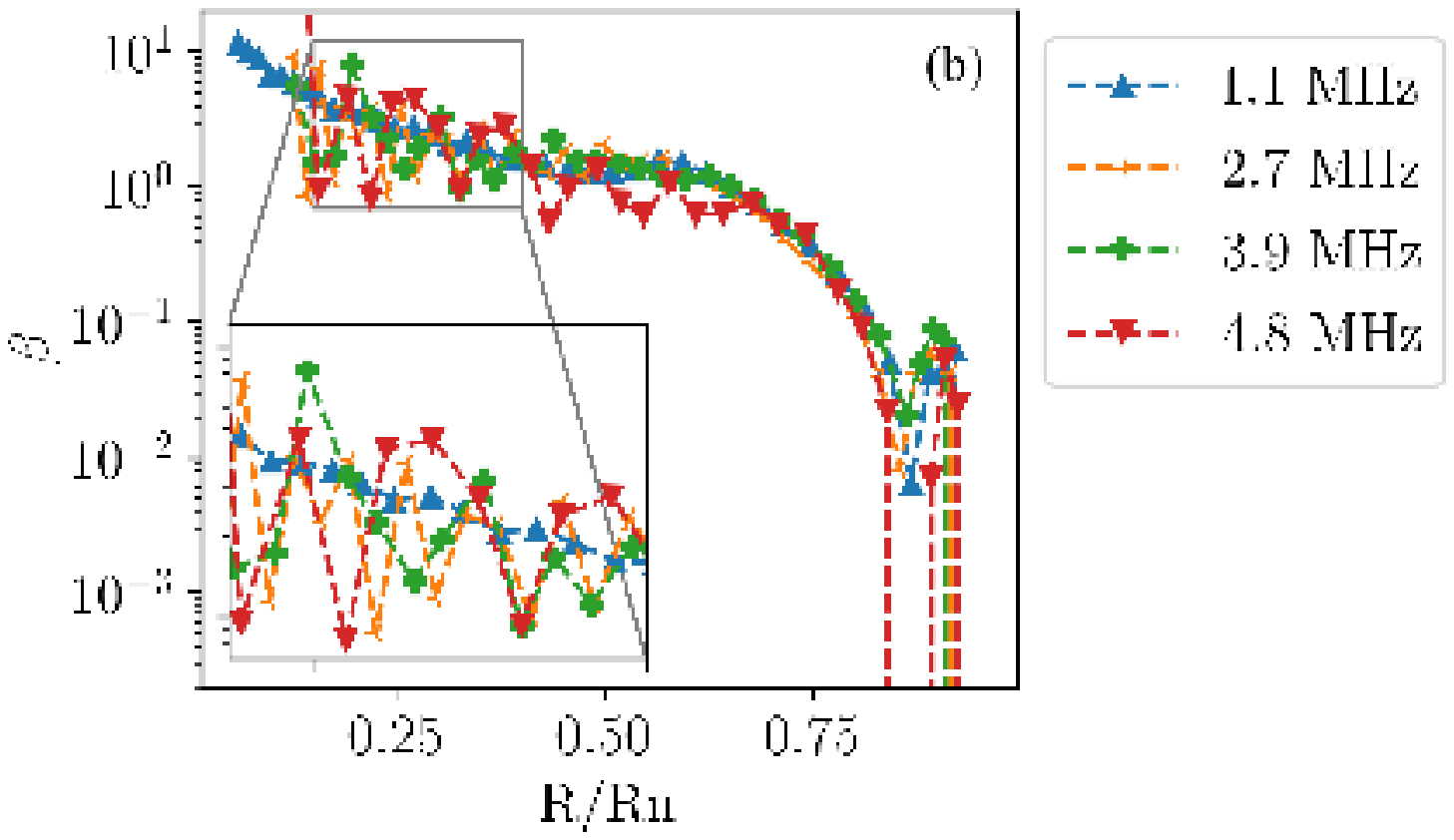}\\% Here is how to import EPS art
\caption{\label{fig:alphabeta} Logarithmic resistance resistivity with respect to temperature $\alpha$ (a) and current $\beta$ (b) as a function of TES working point for some of the pixels in Run 2.}
\end{figure}

\begin{figure}
\includegraphics[width=1.0\linewidth, keepaspectratio]{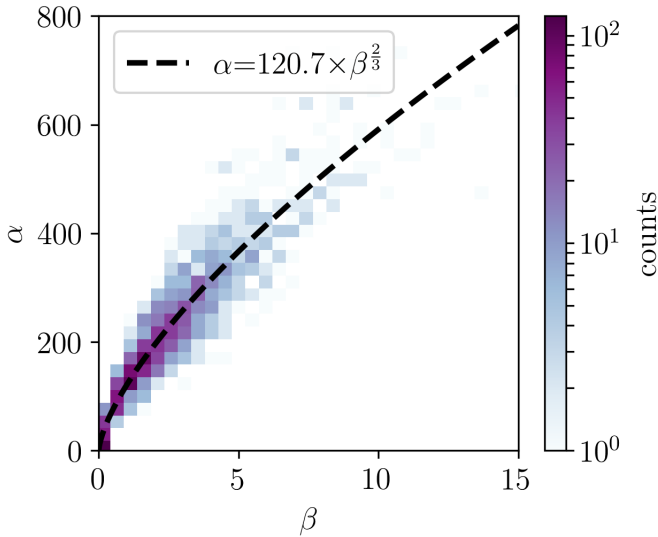}\\% Here is how to import EPS art
\caption{\label{fig:alphavsbeta} $\alpha$ versus $\beta$ for all the 60 pixels that have been measured with empiric relationship between them (solid line).}
\end{figure}

\begin{figure}
\includegraphics[width=1.0\linewidth, keepaspectratio]{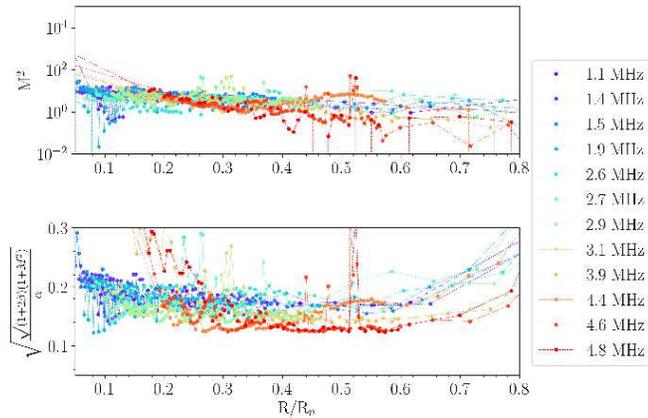}\\% Here is how to import EPS art
\caption{\label{fig:noise} M factor (top plot) and influence of $\alpha$, $\beta$ and $M$ on the energy resolution (lower plot) as a function of the bias point $R/R_\mathrm{n}$ for pixels measured in Run4.}
\end{figure}

Frequency dependent complex impedance measurement $Z(f)$ is a method extensively used to derive the most important thermal and electrical TES parameters\cite{lindeman,taralli3,fukuda2}. All the details about this measurement performed in AC bias has been published in a recent work \cite{taralli}.\\
Data derived from measurements of the frequency dependent complex impedance $Z(f)$  give information about the partial logarithmic derivatives of TES resistance $R(T, I)$ with respect to temperature and current $\alpha$ = $\delta \mathrm{ln}R/\delta \mathrm{ln}T$ and $\beta$ = $\delta \mathrm{ln}R/\delta \mathrm{ln}I$, respectively. In Fig.~\ref{fig:alphabeta}a and Fig.~\ref{fig:alphabeta}b we show $\alpha$ and $\beta$ respectively as a function of $R/R_\mathrm{n}$ for four pixels at different bias frequency. It is worth to notice that, as for the quadrature component of the $IV$ curves (Fig.~\ref{fig:IV_B}b), $\alpha$ and $\beta$ show oscillations as a function of the voltage bias amplitude with a period that depends on the bias frequency as highlighted from the insets (zoom in between 15-40\% of $R/R_\mathrm{n}$) in Fig.~\ref{fig:alphabeta}. The oscillations are large  especially for low values of $R/R_\mathrm{n}$ and have their origin in the earlier mentioned weak-links effect. However, $\alpha$ and $\beta$ are small-signal parameters, and when an X-ray is absorbed by a TES, typically a large part of the transition is sampled, smoothing these peaks. It means that the mean value of  $\alpha$ and $\beta$ at high frequency bias remains close to the one at low bias frequency and the effect on the detector performance is of a secondary importance. The relation between $\alpha$ and $\beta$ is independent of the bias frequency, this is something which becomes clear when $\alpha$ is plotted versus $\beta$. In Fig.~\ref{fig:alphavsbeta} we would like to stress the uniformity of the values of $\alpha$ and $\beta$ among all the pixels in each bias points that have been measured. For instance, the pair $\alpha\sim$200 and $\beta\sim$2.5 appears for more than 40 pixels in that specific bias point (30\% of $R_\mathrm{n}$). The darker pattern highlights the most frequent values among the pixels, while the lighter squares show the values reached by the high frequency pixel during the oscillations. Solid line in Fig.~\ref{fig:alphavsbeta} shows an empiric relationship between $\alpha$ and $\beta$ that could depend on the geometry of the device\cite{martin}.\\
We usually acquire also the detector noise in the same working points through the transition where TES complex impedance measurement has been performed. This turns out to be very important in order to compare the experimental noise with the theoretical one using all the parameters obtained from the fit of the complex impedance. Ideally the detector noise should be explained by means of the phonon noise at low frequencies $S_{P_\mathrm{TFN}} = 4k_\mathrm{B}T^22GF_\mathrm{L}$, the Johnson noise at middle frequencies $S_V = 4k_\mathrm{B}TR(1 + 2\beta)$ and the SQUID or readout noise at high frequencies, where the $T$ an $R$ are the temperature and the resistance of the TES, respectively, $G$ is the thermal conductance to the bath and $F_\mathrm{L}$ is a unitless function that depends on the thermal conductance exponent and on whether phonon transport to the TES is specular or diffuse. The term $1+2\beta$ is the first-order correction to the standard Johnson-noise expression for a non-linear resistor with current dependence\cite{IrwinHilton}. In reality the experimental noise spectrum shows an excess\cite{Ullom} in comparison with the ideal detector calculation in the frequency range where the Johnson noise is dominant. This excess noise is quantified with an additional factor of $(1 + M^2)$ that multiply the expected Johnson noise $S_V = 4k_\mathrm{B}TR(1 + 2\beta)(1 +M^2)$. A number of explanations for this excess Johnson noise have been proposed, however none of the mechanisms gives quantitative predictions consistent with the measured dependencies of the excess electrical noise. At SRON we are currently working to find a full correlation between theoretical and experimental noise in order to understand the magnitude of the unexplained noise specifically on our new high aspect ratio TESs. \\
In Fig.~\ref{fig:noise} we show the value of the fitted $M$ factor for the pixels measured in Run 4 as a function of the bias points (upper plot). At the same time, in the lower plot we can consider together $\alpha$, $\beta$ and $M$ to estimate the effective influence of these parameters on the energy resolution\cite{IrwinHilton} $\Delta E_\mathrm{FWHM} \thickapprox 2.355\sqrt{4k_\mathrm{B} T_\mathrm{0}^2 C\frac{\sqrt{(1 + 2\beta)(1 +M^2)}}{\alpha}}$. This plot shows low bias frequency pixels having almost no oscillations down to 15\% of $R_\mathrm{n}$, whereas shows more oscillating high bias frequency pixels which become more stable around ~35\%-40\% of $R_\mathrm{n}$. This explains why the higher bias frequency pixels need to be biased at 30\%-40\% of $R_\mathrm{n}$ to get the same good energy resolution as the lower bias frequency pixels biased around at 20\%-30\% of the $R_\mathrm{n}$. Of course understanding the physical origins of $\alpha$, $\beta$ and $M$ is essential for further energy resolution optimisation.

\subsection{\label{sec:nep}Noise equivalent power}

\begin{figure}
\includegraphics[width=1\linewidth, keepaspectratio]{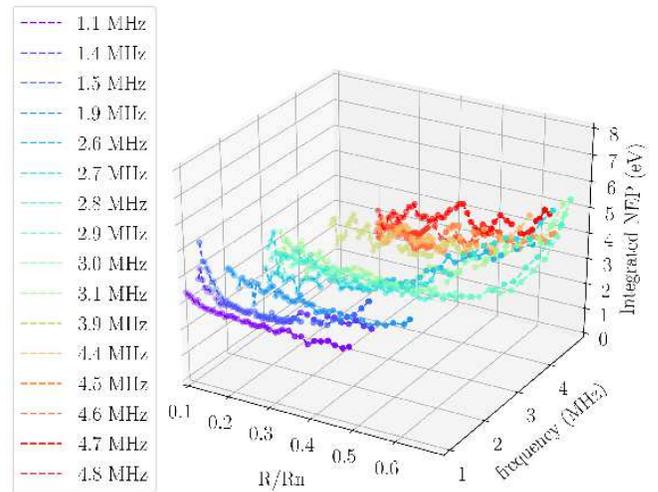}\\
\caption{\label{fig:nep84} 3D-plot of the integrated NEP for each pixel as a function of the working point measured during Run 2.}
\end{figure}

\begin{figure}
\includegraphics[width=1\linewidth, keepaspectratio]{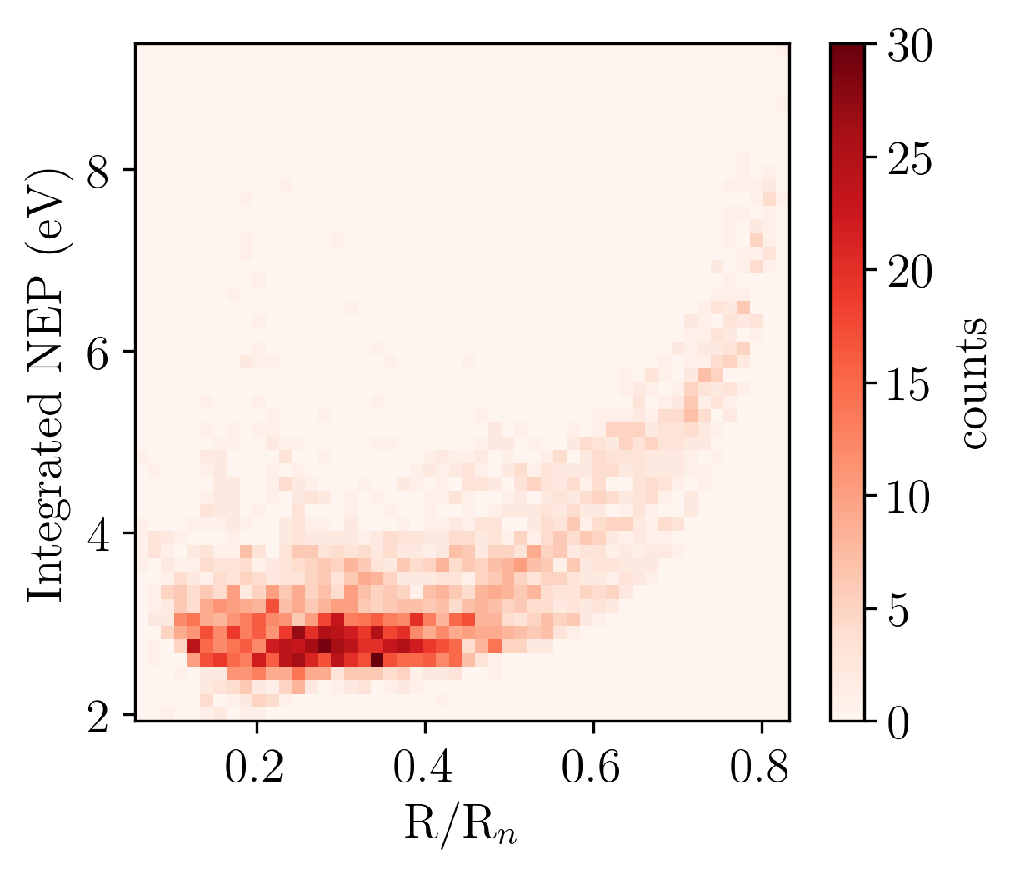}\\% Here is how to import EPS art
\caption{\label{fig:nep2D} Integrated NEP measured for all the 60 pixels during the four Runs as a function of the working point. Most of the pixels show an integrated NEP well below 3 eV between 20 and 40 \% of the normal-state resistance, as clearly highlighted by the red populated area.}
\end{figure}

An $^{55}$Fe X-ray source is placed closely above the array in the set-up and illuminates the entire TES array at a count rate of $\sim$1 cps per pixel for the given absorber with the Mn-K$\alpha$ 5.9 keV fluorescent X-ray line doublet. The noise equivalent power (NEP) has been estimated at different bias points for each pixels by measuring the noise current and the pulse response of the TESs for the connected set of pixels. We typically average 200-300 noise events (with no pulses) and 20 pulse events to calculate the current noise spectra and the FFT of the pulse response. Knowing the energy of the $^{55}$Fe X-ray pulse and dividing the noise spectra by the detector's responsivity, we get the NEP spectra. Integrating the NEP spectra we obtain the baseline resolution that is reported in Fig.~\ref{fig:nep84} for Run 2 as a representative example for all the measured pixels. We observe that at both ends of the transition the average integrated NEP is larger. There is an oscillation as a function of bias resistance which is more prominent for the pixels with higher bias frequency. This behaviour has been already discussed for the quadrature component of $IV$ curves (Fig.~\ref{fig:IV_B}b) and for $\alpha$ and $\beta$ (Fig.~\ref{fig:alphabeta}) being related to the weak-link effect. 
In Fig. ~\ref{fig:nep2D} we plotted all the fobtained integrated NEP values as a function of normalised bias resistance with the aim of predicting the distribution of obtainable energy resolution for the entire array. We can conclude that there is a consistent and populated area from 20\% to 40\% of the normal-state resistance where the integrated NEP is between 2.5 and 3 eV.

\subsection{\label{sec:deltae}Energy resolution}
\begin{figure}
\includegraphics[width=1\linewidth, keepaspectratio]{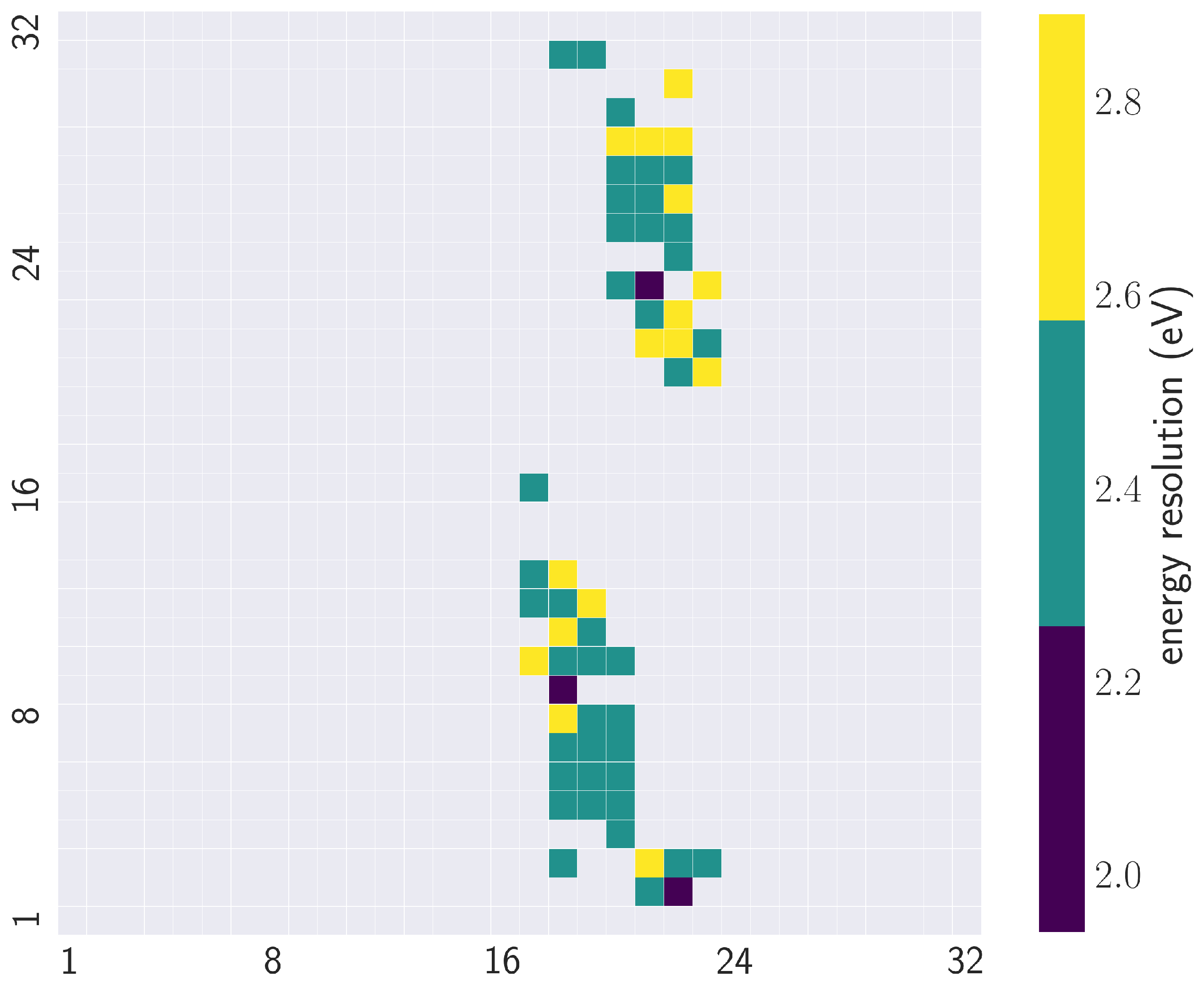}\\% Here is how to import EPS art
\caption{\label{fig:evtotal} Map of the kilo-pixel array X-ray energy resolution considering three different colours: purple (1.98 eV<$\Delta E$<2.28 eV), green (2.28 eV<$\Delta E$<2.58 eV) and yellow (2.58 eV<$\Delta E$<2.88 eV). These ranges reflect the statistical error of $\pm$0.15 eV associated to each measured energy resolution.}
\end{figure}

\begin{figure}
%\subfloat[]{\includegraphics[width=1\linewidth, keepaspectratio]{best_res_all_runs_totalfit.png}\label{fig:evvsfreq}}\\% Here is how to import EPS art
%\subfloat[]{\includegraphics[width=1\linewidth, keepaspectratio]{eV_histo_kilo_pxls.png}\label{fig:evhisto}}\\% Here is how to import EPS art
\includegraphics[width=1\linewidth, keepaspectratio]{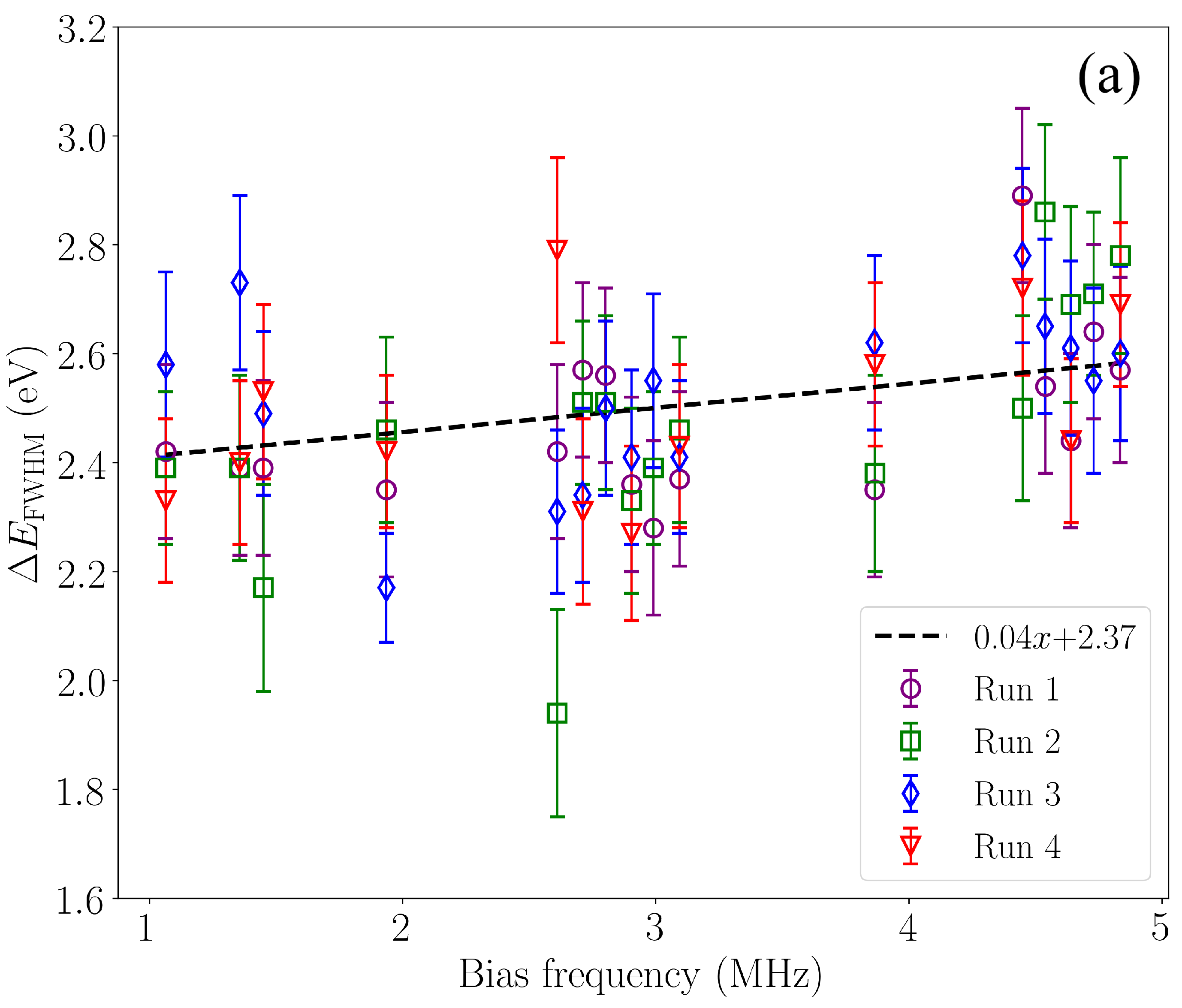}\\% Here is how to import EPS art
%\subfloat[]{\includegraphics[width=1\linewidth, keepaspectratio]{summed_energy.png}\label{fig:holtzer}}\\% Here is how to import EPS art
\includegraphics[width=1\linewidth, keepaspectratio]{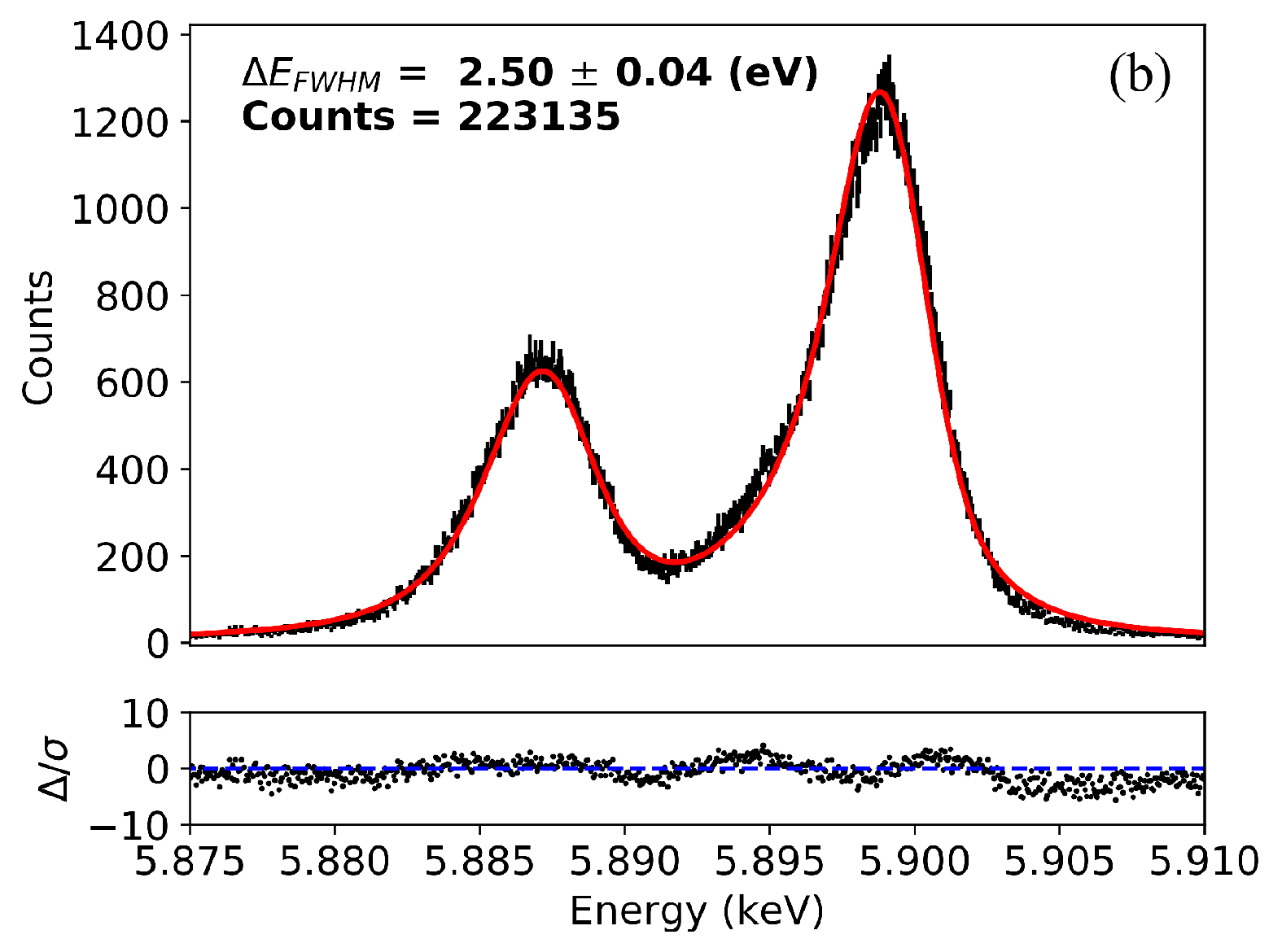}\\% Here is how to import EPS art
\caption{\label{fig:resvsfreq}(a) Typical energy resolution measured at the best bias points as a function of the pixel's bias frequency for the four Runs (dots) and corresponding linear fit to evaluate degradation between low and high bias frequency (dashed line). (b) Summed X-ray spectra at 5.9 keV over the 60 pixels measured at the best bias points during the four Runs. The solid line is the best fit to the data, the points are the measured Mn-K$\alpha$ emission lines and the dashed line is the intrinsic emission of the source. The lower plot shows the residuals of the fit normalized by the error bars.}
\end{figure}

Integrating the measured NEP over a wide bias range identifies the promising working points where good energy resolution can be expected. In Fig.~\ref{fig:evtotal} we summarise the best energy resolution for each pixel that have been measured from one side of the array to the other. We usually collect around 5000 X-Ray photons to get a statistical error of about $\pm$0.15 eV for the reported energy resolution. Although two (or more) pixels show different absolute values of their energy resolution, they can be considered equivalent inside the statistical error. We can generate a map of the single pixel performance dividing the measured energy resolutions in three different ranges (different colors in Fig.~\ref{fig:evtotal}) separated by the statistical error associated to each measurement ($\pm$0.15 eV). We provide an immediate picture where the majority of the pixels show an energy resolution between 2.28 eV and 2.58 eV.
We can also consider any dependence of the energy resolution on the bias frequency of the pixel under test. Fig.~\ref{fig:resvsfreq}a shows the typical energy resolution measured at the best bias points as a function of the pixel's bias frequency for the four Runs. From the linear fit (dashed line)  appears visible a degradation of the energy resolution between low and high bias frequency pixels of about 4\%, which is considerably less than the 14\% reported with the previous pixel design\cite{fdm1}. This confirms the effectiveness of the new pixels design and the relative efforts to reduce the undesired effects depending on the bias frequency, achieving in this way global and uniform good performances as highlighted by a summed X-ray energy resolution of 2.50$\pm$0.04 eV. Fig.~\ref{fig:resvsfreq}b shows the summed $^{55}$Fe X-Ray spectra over the 60 pixels measured at the best bias points during the four Runs that is representative of the quality of the whole array.

\section{\label{sec:conclusion}Conclusion}

We have extensively characterised our 32$\times$32 pixel array by measuring 60 pixels concentrated in two different quadrants of the chip. The goal of this study was to evaluate the uniformity of such large array in terms of thermal and electrical parameters, noise equivalent power and energy resolution.\\
We found a mean critical temperature $T_\mathrm{c}$ = 89.5 with a standard deviation of 0.5 mK. The temperature variation within the same quadrant is less than 0.6 mK and less than 1.5 mK across the array ($\sim$1 cm). The thermal conductance $G$ has been found to be 117$\pm$17 pW/K, which is larger than expected. It turned out that the stems used to sustain and connect the absorber to the TES play a role in its final value. A further reducing of the stems diameter as well as the number of the stems directly connected to the membrane will reduce this additional thermal conductance.\\
We obtained uniform values of $\alpha$, $\beta$ and $M$ factor all over the measured pixels. Values of  $\alpha \sim$ 200 and $\beta \sim$ 2.5 have been found between the 20\% and 40\% of the normal-state resistance for pixels connected to lower bias frequencies. Same averaged values but with oscillations (due to the weak-links effect) can be found for the pixels connected to higher frequencies. The excess noise expressed by the $M$ factor, shows consistency over a wide range of working points and all over the measured pixels .\\
A summed energy resolution of 2.50$\pm$0.04 eV at 5.9 keV has been evaluated over the 60 pixels. 
This large array has shown broad uniformity and high performances, and is therefore very promising to offer technology for future X-ray space missions. The entire fabrication process has been optimized to get arrays suitable for FDM readout which means low resistance per square and high aspect-ratio pixels to reduce Josephson effects that affect the performance of a TES microcalorimeter under AC bias. Different pixel design and analogous fabrication process can lead to optimized pixels suitable for time domain multiplexing readout intended to meet the detector requirements of
the X-IFU instrument.

\begin{acknowledgments}
This work is partly funded by European Space Agency (ESA) and coordinated with other European efforts under ESA CTP contract ITT AO/1-7947/14/NL/BW. It has also received funding from the European Union's Horizon 2020 Programme under the AHEAD (Activities for the High-Energy Astrophysics Domain) project with grant agreement number 654215.
\end{acknowledgments}

\section*{data availability}
The data that support the findings of this study are available from the corresponding author upon reasonable request.
%\end{data availability}

%\nocite{*}

\bibliography{article}% Produces the bibliography via BibTeX.

\providecommand{\noopsort}[1]{}\providecommand{\singleletter}[1]{#1}%
\begin{thebibliography}{37}%
\makeatletter
\providecommand \@ifxundefined [1]{%
 \@ifx{#1\undefined}
}%
\providecommand \@ifnum [1]{%
 \ifnum #1\expandafter \@firstoftwo
 \else \expandafter \@secondoftwo
 \fi
}%
\providecommand \@ifx [1]{%
 \ifx #1\expandafter \@firstoftwo
 \else \expandafter \@secondoftwo
 \fi
}%
\providecommand \natexlab [1]{#1}%
\providecommand \enquote  [1]{``#1''}%
\providecommand \bibnamefont  [1]{#1}%
\providecommand \bibfnamefont [1]{#1}%
\providecommand \citenamefont [1]{#1}%
\providecommand \href@noop [0]{\@secondoftwo}%
\providecommand \href [0]{\begingroup \@sanitize@url \@href}%
\providecommand \@href[1]{\@@startlink{#1}\@@href}%
\providecommand \@@href[1]{\endgroup#1\@@endlink}%
\providecommand \@sanitize@url [0]{\catcode `\\12\catcode `\$12\catcode
  `\&12\catcode `\#12\catcode `\^12\catcode `\_12\catcode `\%12\relax}%
\providecommand \@@startlink[1]{}%
\providecommand \@@endlink[0]{}%
\providecommand \url  [0]{\begingroup\@sanitize@url \@url }%
\providecommand \@url [1]{\endgroup\@href {#1}{\urlprefix }}%
\providecommand \urlprefix  [0]{URL }%
\providecommand \Eprint [0]{\href }%
\providecommand \doibase [0]{http://dx.doi.org/}%
\providecommand \selectlanguage [0]{\@gobble}%
\providecommand \bibinfo  [0]{\@secondoftwo}%
\providecommand \bibfield  [0]{\@secondoftwo}%
\providecommand \translation [1]{[#1]}%
\providecommand \BibitemOpen [0]{}%
\providecommand \bibitemStop [0]{}%
\providecommand \bibitemNoStop [0]{.\EOS\space}%
\providecommand \EOS [0]{\spacefactor3000\relax}%
\providecommand \BibitemShut  [1]{\csname bibitem#1\endcsname}%
\let\auto@bib@innerbib\@empty
%</preamble>
\bibitem [{\citenamefont {Irwin}\ and\ \citenamefont
  {Hilton}(2005)}]{IrwinHilton}%
  \BibitemOpen
  \bibfield  {author} {\bibinfo {author} {\bibfnamefont {K.~D.}\ \bibnamefont
  {Irwin}}\ and\ \bibinfo {author} {\bibfnamefont {G.~C.~W.}\ \bibnamefont
  {Hilton}},\ }\enquote {\bibinfo {title} {Cryogenic particle detection. topics
  in applied physics},}\ \ (\bibinfo  {publisher} {Springer, Berlin,
  Heidelberg},\ \bibinfo {year} {2005})\ Chap.\ \bibinfo {chapter}
  {Transition-Edge Sensors}, pp.\ \bibinfo {pages} {63--152}\BibitemShut
  {NoStop}%
\bibitem [{\citenamefont {Noroozian}\ \emph {et~al.}(2013)\citenamefont
  {Noroozian}, \citenamefont {Mates}, \citenamefont {Bennett}, \citenamefont
  {Brevik}, \citenamefont {Fowler}, \citenamefont {Gao}, \citenamefont
  {Hilton}, \citenamefont {Horansky}, \citenamefont {Irwin}, \citenamefont
  {Kang}, \citenamefont {Schmidt}, \citenamefont {Vale},\ and\ \citenamefont
  {Ullom}}]{Ullom2}%
  \BibitemOpen
  \bibfield  {author} {\bibinfo {author} {\bibfnamefont {O.}~\bibnamefont
  {Noroozian}}, \bibinfo {author} {\bibfnamefont {J.~A.~B.}\ \bibnamefont
  {Mates}}, \bibinfo {author} {\bibfnamefont {D.~A.}\ \bibnamefont {Bennett}},
  \bibinfo {author} {\bibfnamefont {J.~A.}\ \bibnamefont {Brevik}}, \bibinfo
  {author} {\bibfnamefont {J.~W.}\ \bibnamefont {Fowler}}, \bibinfo {author}
  {\bibfnamefont {J.}~\bibnamefont {Gao}}, \bibinfo {author} {\bibfnamefont
  {G.~C.}\ \bibnamefont {Hilton}}, \bibinfo {author} {\bibfnamefont {R.~D.}\
  \bibnamefont {Horansky}}, \bibinfo {author} {\bibfnamefont {K.~D.}\
  \bibnamefont {Irwin}}, \bibinfo {author} {\bibfnamefont {Z.}~\bibnamefont
  {Kang}}, \bibinfo {author} {\bibfnamefont {D.~R.}\ \bibnamefont {Schmidt}},
  \bibinfo {author} {\bibfnamefont {L.~R.}\ \bibnamefont {Vale}}, \ and\
  \bibinfo {author} {\bibfnamefont {J.~N.}\ \bibnamefont {Ullom}},\ }\href@noop
  {} {\bibfield  {journal} {\bibinfo  {journal} {Appl. Phys. Lett.}\ }\textbf
  {\bibinfo {volume} {103}},\ \bibinfo {pages} {202602} (\bibinfo {year}
  {2013})}\BibitemShut {NoStop}%
\bibitem [{\citenamefont {Ullom}\ \emph {et~al.}(2005)\citenamefont {Ullom},
  \citenamefont {Beall}, \citenamefont {Doriese}, \citenamefont {Duncan},
  \citenamefont {Ferreira}, \citenamefont {Hilton}, \citenamefont {Irwin},
  \citenamefont {Reintsema},\ and\ \citenamefont {Vale}}]{Ullom}%
  \BibitemOpen
  \bibfield  {author} {\bibinfo {author} {\bibfnamefont {J.~N.}\ \bibnamefont
  {Ullom}}, \bibinfo {author} {\bibfnamefont {J.~A.}\ \bibnamefont {Beall}},
  \bibinfo {author} {\bibfnamefont {W.~B.}\ \bibnamefont {Doriese}}, \bibinfo
  {author} {\bibfnamefont {W.~D.}\ \bibnamefont {Duncan}}, \bibinfo {author}
  {\bibfnamefont {L.}~\bibnamefont {Ferreira}}, \bibinfo {author}
  {\bibfnamefont {G.~C.}\ \bibnamefont {Hilton}}, \bibinfo {author}
  {\bibfnamefont {K.~D.}\ \bibnamefont {Irwin}}, \bibinfo {author}
  {\bibfnamefont {C.~D.}\ \bibnamefont {Reintsema}}, \ and\ \bibinfo {author}
  {\bibfnamefont {L.~R.}\ \bibnamefont {Vale}},\ }\href@noop {} {\bibfield
  {journal} {\bibinfo  {journal} {Appl. Phys. Lett.}\ }\textbf {\bibinfo
  {volume} {87}},\ \bibinfo {pages} {194103} (\bibinfo {year}
  {2005})}\BibitemShut {NoStop}%
\bibitem [{\citenamefont {Miniussi}\ \emph {et~al.}(2018)\citenamefont
  {Miniussi}, \citenamefont {Adams}, \citenamefont {Bandler}, \citenamefont
  {Chervenak}, \citenamefont {Datesman}, \citenamefont {Eckart}, \citenamefont
  {Ewin}, \citenamefont {Finkbeiner}, \citenamefont {Kelley}, \citenamefont
  {Kilbourne}, \citenamefont {Porter}, \citenamefont {Sadleir}, \citenamefont
  {Sakai}, \citenamefont {Smith}, \citenamefont {Wakeham}, \citenamefont
  {Wassell},\ and\ \citenamefont {Yoon}}]{minussi}%
  \BibitemOpen
  \bibfield  {author} {\bibinfo {author} {\bibfnamefont {A.~R.}\ \bibnamefont
  {Miniussi}}, \bibinfo {author} {\bibfnamefont {J.~S.}\ \bibnamefont {Adams}},
  \bibinfo {author} {\bibfnamefont {S.~R.}\ \bibnamefont {Bandler}}, \bibinfo
  {author} {\bibfnamefont {J.~A.}\ \bibnamefont {Chervenak}}, \bibinfo {author}
  {\bibfnamefont {A.~M.}\ \bibnamefont {Datesman}}, \bibinfo {author}
  {\bibfnamefont {M.~E.}\ \bibnamefont {Eckart}}, \bibinfo {author}
  {\bibfnamefont {A.~J.}\ \bibnamefont {Ewin}}, \bibinfo {author}
  {\bibfnamefont {F.~M.}\ \bibnamefont {Finkbeiner}}, \bibinfo {author}
  {\bibfnamefont {R.~L.}\ \bibnamefont {Kelley}}, \bibinfo {author}
  {\bibfnamefont {C.~A.}\ \bibnamefont {Kilbourne}}, \bibinfo {author}
  {\bibfnamefont {F.~S.}\ \bibnamefont {Porter}}, \bibinfo {author}
  {\bibfnamefont {J.~E.}\ \bibnamefont {Sadleir}}, \bibinfo {author}
  {\bibfnamefont {K.}~\bibnamefont {Sakai}}, \bibinfo {author} {\bibfnamefont
  {S.~J.}\ \bibnamefont {Smith}}, \bibinfo {author} {\bibfnamefont {N.~A.}\
  \bibnamefont {Wakeham}}, \bibinfo {author} {\bibfnamefont {E.~J.}\
  \bibnamefont {Wassell}}, \ and\ \bibinfo {author} {\bibfnamefont
  {W.}~\bibnamefont {Yoon}},\ }\href@noop {} {\bibfield  {journal} {\bibinfo
  {journal} {J. Low Temp. Phys.}\ }\textbf {\bibinfo {volume} {193}},\ \bibinfo
  {pages} {337} (\bibinfo {year} {2018})}\BibitemShut {NoStop}%
\bibitem [{\citenamefont {Brink}(2012)}]{brink}%
  \BibitemOpen
  \bibfield  {author} {\bibinfo {author} {\bibfnamefont {P.~L.}\ \bibnamefont
  {Brink}},\ }\href@noop {} {\bibfield  {journal} {\bibinfo  {journal} {J. Low
  Temp. Phys.}\ }\textbf {\bibinfo {volume} {167}},\ \bibinfo {pages} {1048}
  (\bibinfo {year} {2012})}\BibitemShut {NoStop}%
\bibitem [{\citenamefont {Lolli}\ \emph {et~al.}(2013)\citenamefont {Lolli},
  \citenamefont {Taralli}, \citenamefont {Portesi}, \citenamefont {Rajteri},\
  and\ \citenamefont {Monticone}}]{lolli}%
  \BibitemOpen
  \bibfield  {author} {\bibinfo {author} {\bibfnamefont {L.}~\bibnamefont
  {Lolli}}, \bibinfo {author} {\bibfnamefont {E.}~\bibnamefont {Taralli}},
  \bibinfo {author} {\bibfnamefont {C.}~\bibnamefont {Portesi}}, \bibinfo
  {author} {\bibfnamefont {M.}~\bibnamefont {Rajteri}}, \ and\ \bibinfo
  {author} {\bibfnamefont {E.}~\bibnamefont {Monticone}},\ }\href@noop {}
  {\bibfield  {journal} {\bibinfo  {journal} {Appl. Phys. Lett.}\ }\textbf
  {\bibinfo {volume} {103}},\ \bibinfo {pages} {041107} (\bibinfo {year}
  {2013})}\BibitemShut {NoStop}%
\bibitem [{\citenamefont {Cabrera}\ \emph {et~al.}(1998)\citenamefont
  {Cabrera}, \citenamefont {Clarke}, \citenamefont {Colling}, \citenamefont
  {Miller}, \citenamefont {Nam},\ and\ \citenamefont {Romani}}]{cabrera}%
  \BibitemOpen
  \bibfield  {author} {\bibinfo {author} {\bibfnamefont {B.}~\bibnamefont
  {Cabrera}}, \bibinfo {author} {\bibfnamefont {R.~M.}\ \bibnamefont {Clarke}},
  \bibinfo {author} {\bibfnamefont {P.}~\bibnamefont {Colling}}, \bibinfo
  {author} {\bibfnamefont {A.~J.}\ \bibnamefont {Miller}}, \bibinfo {author}
  {\bibfnamefont {S.}~\bibnamefont {Nam}}, \ and\ \bibinfo {author}
  {\bibfnamefont {R.~W.}\ \bibnamefont {Romani}},\ }\href@noop {} {\bibfield
  {journal} {\bibinfo  {journal} {Appl. Phys. Lett.}\ }\textbf {\bibinfo
  {volume} {73}},\ \bibinfo {pages} {735} (\bibinfo {year} {1998})}\BibitemShut
  {NoStop}%
\bibitem [{\citenamefont {Niwa}\ \emph {et~al.}(2017)\citenamefont {Niwa},
  \citenamefont {Numata}, \citenamefont {Hattori},\ and\ \citenamefont
  {Fukuda}}]{fukuda}%
  \BibitemOpen
  \bibfield  {author} {\bibinfo {author} {\bibfnamefont {K.}~\bibnamefont
  {Niwa}}, \bibinfo {author} {\bibfnamefont {T.}~\bibnamefont {Numata}},
  \bibinfo {author} {\bibfnamefont {K.}~\bibnamefont {Hattori}}, \ and\
  \bibinfo {author} {\bibfnamefont {D.}~\bibnamefont {Fukuda}},\ }\href@noop {}
  {\bibfield  {journal} {\bibinfo  {journal} {Sci. Rep.}\ }\textbf {\bibinfo
  {volume} {7}},\ \bibinfo {pages} {45660} (\bibinfo {year}
  {2017})}\BibitemShut {NoStop}%
\bibitem [{\citenamefont {O'Brient}\ \emph {et~al.}(2013)\citenamefont
  {O'Brient}, \citenamefont {Ade}, \citenamefont {Arnold}, \citenamefont
  {Edwards}, \citenamefont {Engargiola}, \citenamefont {Holzapfel},
  \citenamefont {Lee}, \citenamefont {Myers}, \citenamefont {Quealy},
  \citenamefont {Rebeiz}, \citenamefont {Richards},\ and\ \citenamefont
  {Suzuki}}]{obrient}%
  \BibitemOpen
  \bibfield  {author} {\bibinfo {author} {\bibfnamefont {R.}~\bibnamefont
  {O'Brient}}, \bibinfo {author} {\bibfnamefont {P.}~\bibnamefont {Ade}},
  \bibinfo {author} {\bibfnamefont {K.}~\bibnamefont {Arnold}}, \bibinfo
  {author} {\bibfnamefont {J.}~\bibnamefont {Edwards}}, \bibinfo {author}
  {\bibfnamefont {G.}~\bibnamefont {Engargiola}}, \bibinfo {author}
  {\bibfnamefont {W.~L.}\ \bibnamefont {Holzapfel}}, \bibinfo {author}
  {\bibfnamefont {A.~T.}\ \bibnamefont {Lee}}, \bibinfo {author} {\bibfnamefont
  {M.~J.}\ \bibnamefont {Myers}}, \bibinfo {author} {\bibfnamefont
  {E.}~\bibnamefont {Quealy}}, \bibinfo {author} {\bibfnamefont
  {G.}~\bibnamefont {Rebeiz}}, \bibinfo {author} {\bibfnamefont
  {P.}~\bibnamefont {Richards}}, \ and\ \bibinfo {author} {\bibfnamefont
  {A.}~\bibnamefont {Suzuki}},\ }\href@noop {} {\bibfield  {journal} {\bibinfo
  {journal} {Appl. Phys. Lett.}\ }\textbf {\bibinfo {volume} {102}},\ \bibinfo
  {pages} {063506} (\bibinfo {year} {2013})}\BibitemShut {NoStop}%
\bibitem [{\citenamefont {Li}\ \emph {et~al.}(2018)\citenamefont {Li},
  \citenamefont {Alpert}, \citenamefont {Becker}, \citenamefont {Bennett},
  \citenamefont {Carini}, \citenamefont {Cho}, \citenamefont {Doriese},
  \citenamefont {Dusatko}, \citenamefont {Fowler}, \citenamefont {Frisch},
  \citenamefont {Gard}, \citenamefont {Guillet}, \citenamefont {Hilton},
  \citenamefont {Holmes}, \citenamefont {Irwin}, \citenamefont {Kotsubo},
  \citenamefont {Lee}, \citenamefont {Mates}, \citenamefont {Morgan},
  \citenamefont {Nakahara}, \citenamefont {Pappas}, \citenamefont {Reintsema},
  \citenamefont {Schmidt}, \citenamefont {Smith}, \citenamefont {Swetz},
  \citenamefont {Thayer}, \citenamefont {Titus}, \citenamefont {Ullom},
  \citenamefont {Vale}, \citenamefont {Winkle}, \citenamefont {Wessels},\ and\
  \citenamefont {Zhang}}]{li}%
  \BibitemOpen
  \bibfield  {author} {\bibinfo {author} {\bibfnamefont {D.}~\bibnamefont
  {Li}}, \bibinfo {author} {\bibfnamefont {B.~K.}\ \bibnamefont {Alpert}},
  \bibinfo {author} {\bibfnamefont {D.~T.}\ \bibnamefont {Becker}}, \bibinfo
  {author} {\bibfnamefont {D.~A.}\ \bibnamefont {Bennett}}, \bibinfo {author}
  {\bibfnamefont {G.~A.}\ \bibnamefont {Carini}}, \bibinfo {author}
  {\bibfnamefont {H.-M.}\ \bibnamefont {Cho}}, \bibinfo {author} {\bibfnamefont
  {W.~B.}\ \bibnamefont {Doriese}}, \bibinfo {author} {\bibfnamefont {J.~E.}\
  \bibnamefont {Dusatko}}, \bibinfo {author} {\bibfnamefont {J.~W.}\
  \bibnamefont {Fowler}}, \bibinfo {author} {\bibfnamefont {J.~C.}\
  \bibnamefont {Frisch}}, \bibinfo {author} {\bibfnamefont {J.~D.}\
  \bibnamefont {Gard}}, \bibinfo {author} {\bibfnamefont {S.}~\bibnamefont
  {Guillet}}, \bibinfo {author} {\bibfnamefont {G.~C.}\ \bibnamefont {Hilton}},
  \bibinfo {author} {\bibfnamefont {M.~R.}\ \bibnamefont {Holmes}}, \bibinfo
  {author} {\bibfnamefont {K.~D.}\ \bibnamefont {Irwin}}, \bibinfo {author}
  {\bibfnamefont {V.}~\bibnamefont {Kotsubo}}, \bibinfo {author} {\bibfnamefont
  {S.-J.}\ \bibnamefont {Lee}}, \bibinfo {author} {\bibfnamefont {J.~A.~B.}\
  \bibnamefont {Mates}}, \bibinfo {author} {\bibfnamefont {K.~M.}\ \bibnamefont
  {Morgan}}, \bibinfo {author} {\bibfnamefont {K.}~\bibnamefont {Nakahara}},
  \bibinfo {author} {\bibfnamefont {C.~G.}\ \bibnamefont {Pappas}}, \bibinfo
  {author} {\bibfnamefont {C.~D.}\ \bibnamefont {Reintsema}}, \bibinfo {author}
  {\bibfnamefont {D.~R.}\ \bibnamefont {Schmidt}}, \bibinfo {author}
  {\bibfnamefont {S.~R.}\ \bibnamefont {Smith}}, \bibinfo {author}
  {\bibfnamefont {D.~S.}\ \bibnamefont {Swetz}}, \bibinfo {author}
  {\bibfnamefont {J.~B.}\ \bibnamefont {Thayer}}, \bibinfo {author}
  {\bibfnamefont {C.~J.}\ \bibnamefont {Titus}}, \bibinfo {author}
  {\bibfnamefont {J.~N.}\ \bibnamefont {Ullom}}, \bibinfo {author}
  {\bibfnamefont {L.~R.}\ \bibnamefont {Vale}}, \bibinfo {author}
  {\bibfnamefont {D.~D.~V.}\ \bibnamefont {Winkle}}, \bibinfo {author}
  {\bibfnamefont {A.}~\bibnamefont {Wessels}}, \ and\ \bibinfo {author}
  {\bibfnamefont {L.}~\bibnamefont {Zhang}},\ }\href@noop {} {\bibfield
  {journal} {\bibinfo  {journal} {J. Low Temp. Phys.}\ }\textbf {\bibinfo
  {volume} {193}},\ \bibinfo {pages} {1287} (\bibinfo {year}
  {2018})}\BibitemShut {NoStop}%
\bibitem [{\citenamefont {{SPICA collaboration}}(2018)}]{spica}%
  \BibitemOpen
  \bibfield  {author} {\bibinfo {author} {\bibnamefont {{SPICA
  collaboration}}},\ }\href@noop {} {\bibfield  {journal} {\bibinfo  {journal}
  {Publications of the Astronomical Society of Australia (PASA)}\ }\textbf
  {\bibinfo {volume} {35}},\ \bibinfo {pages} {E030} (\bibinfo {year}
  {2018})}\BibitemShut {NoStop}%
\bibitem [{\citenamefont {Barret}\ \emph {et~al.}(2019)\citenamefont {Barret},
  \citenamefont {Decourchelle}, \citenamefont {Fabian}, \citenamefont
  {Guainazzi}, \citenamefont {Nandra}, \citenamefont {Smith},\ and\
  \citenamefont {den Herder}}]{athena}%
  \BibitemOpen
  \bibfield  {author} {\bibinfo {author} {\bibfnamefont {D.}~\bibnamefont
  {Barret}}, \bibinfo {author} {\bibfnamefont {A.}~\bibnamefont
  {Decourchelle}}, \bibinfo {author} {\bibfnamefont {A.}~\bibnamefont
  {Fabian}}, \bibinfo {author} {\bibfnamefont {M.}~\bibnamefont {Guainazzi}},
  \bibinfo {author} {\bibfnamefont {K.}~\bibnamefont {Nandra}}, \bibinfo
  {author} {\bibfnamefont {R.}~\bibnamefont {Smith}}, \ and\ \bibinfo {author}
  {\bibfnamefont {J.-W.}\ \bibnamefont {den Herder}},\ }\href@noop {} {\enquote
  {\bibinfo {title} {{The Athena space X-ray Observatory and the astrophysics
  of hot plasma}},}\ }\bibinfo {howpublished} {arXiv:1912.04615 [astro-ph.IM]}
  (\bibinfo {year} {2019})\BibitemShut {NoStop}%
\bibitem [{\citenamefont {{HUBS collaboration}}(2020)}]{hubs}%
  \BibitemOpen
  \bibfield  {author} {\bibinfo {author} {\bibnamefont {{HUBS
  collaboration}}},\ }\href@noop {} {}\bibinfo {howpublished}
  {http://hubs.phys.tsinghua.edu.cn/en/collaboration.html} (\bibinfo {year}
  {2020}),\ \bibinfo {note} {website}\BibitemShut {NoStop}%
\bibitem [{\citenamefont {{XIFU collaboration}}()}]{xifu}%
  \BibitemOpen
  \bibfield  {author} {\bibinfo {author} {\bibnamefont {{XIFU
  collaboration}}},\ }\href@noop {} {\enquote {\bibinfo {title} {{The ATHENA
  X-ray Integral Field Unit (X-IFU)}},}\ }\bibinfo {howpublished} {Proc. SPIE
  10699, Space Telescopes and Instrumentation 2018: Ultraviolet to Gamma Ray,
  106991G (31 July 2018)}\BibitemShut {NoStop}%
\bibitem [{\citenamefont {Akamatsu}\ \emph {et~al.}(2020)\citenamefont
  {Akamatsu}, \citenamefont {Gottardi}, \citenamefont {van~der Kuur},
  \citenamefont {de~Vries}, \citenamefont {Bruijn}, \citenamefont {Chervenak},
  \citenamefont {Kiviranta}, \citenamefont {van~den Linden}, \citenamefont
  {andA. Miniussi}, \citenamefont {Ravensberg}, \citenamefont {Sakai},
  \citenamefont {Smith},\ and\ \citenamefont {Wakeham}}]{fdm1}%
  \BibitemOpen
  \bibfield  {author} {\bibinfo {author} {\bibfnamefont {H.}~\bibnamefont
  {Akamatsu}}, \bibinfo {author} {\bibfnamefont {L.}~\bibnamefont {Gottardi}},
  \bibinfo {author} {\bibfnamefont {J.}~\bibnamefont {van~der Kuur}}, \bibinfo
  {author} {\bibfnamefont {C.~P.}\ \bibnamefont {de~Vries}}, \bibinfo {author}
  {\bibfnamefont {M.~P.}\ \bibnamefont {Bruijn}}, \bibinfo {author}
  {\bibfnamefont {J.~A.}\ \bibnamefont {Chervenak}}, \bibinfo {author}
  {\bibfnamefont {M.}~\bibnamefont {Kiviranta}}, \bibinfo {author}
  {\bibfnamefont {A.~J.}\ \bibnamefont {van~den Linden}}, \bibinfo {author}
  {\bibfnamefont {B.~D.~J.}\ \bibnamefont {andA. Miniussi}}, \bibinfo {author}
  {\bibfnamefont {K.}~\bibnamefont {Ravensberg}}, \bibinfo {author}
  {\bibfnamefont {K.}~\bibnamefont {Sakai}}, \bibinfo {author} {\bibfnamefont
  {S.~J.}\ \bibnamefont {Smith}}, \ and\ \bibinfo {author} {\bibfnamefont
  {N.}~\bibnamefont {Wakeham}},\ }\href@noop {} {\bibfield  {journal} {\bibinfo
   {journal} {J. Low Temp. Phys.}\ }\textbf {\bibinfo {volume} {199}},\
  \bibinfo {pages} {737} (\bibinfo {year} {2020})}\BibitemShut {NoStop}%
\bibitem [{\citenamefont {Sakai}\ \emph {et~al.}(2016)\citenamefont {Sakai},
  \citenamefont {Yamamoto}, \citenamefont {Takei}, \citenamefont {Mitsuda},
  \citenamefont {Yamasaki}, \citenamefont {Hidaka}, \citenamefont {Nagasawa},
  \citenamefont {Kohjiro},\ and\ \citenamefont {Miyazaki}}]{fdm2}%
  \BibitemOpen
  \bibfield  {author} {\bibinfo {author} {\bibfnamefont {K.}~\bibnamefont
  {Sakai}}, \bibinfo {author} {\bibfnamefont {R.}~\bibnamefont {Yamamoto}},
  \bibinfo {author} {\bibfnamefont {Y.}~\bibnamefont {Takei}}, \bibinfo
  {author} {\bibfnamefont {K.}~\bibnamefont {Mitsuda}}, \bibinfo {author}
  {\bibfnamefont {N.~Y.}\ \bibnamefont {Yamasaki}}, \bibinfo {author}
  {\bibfnamefont {M.}~\bibnamefont {Hidaka}}, \bibinfo {author} {\bibfnamefont
  {S.}~\bibnamefont {Nagasawa}}, \bibinfo {author} {\bibfnamefont
  {S.}~\bibnamefont {Kohjiro}}, \ and\ \bibinfo {author} {\bibfnamefont
  {T.}~\bibnamefont {Miyazaki}},\ }\href@noop {} {\bibfield  {journal}
  {\bibinfo  {journal} {J. Low Temp. Phys.}\ }\textbf {\bibinfo {volume}
  {184}},\ \bibinfo {pages} {519} (\bibinfo {year} {2016})}\BibitemShut
  {NoStop}%
\bibitem [{\citenamefont {Yoon}\ \emph {et~al.}(2018)\citenamefont {Yoon},
  \citenamefont {Adams}, \citenamefont {Bandler}, \citenamefont {Becker},
  \citenamefont {Bennett4}, \citenamefont {Chervenak}, \citenamefont
  {Datesman}, \citenamefont {Eckart}, \citenamefont {Finkbeiner}, \citenamefont
  {Fowler}, \citenamefont {Gard}, \citenamefont {Hilton}, \citenamefont
  {Kelley}, \citenamefont {Kilbourne}, \citenamefont {Mates}, \citenamefont
  {Miniussi}, \citenamefont {Moseley}, \citenamefont {Noroozian}, \citenamefont
  {Porter}, \citenamefont {Reintsema}, \citenamefont {Sadleir}, \citenamefont
  {Sakai}, \citenamefont {Smith}, \citenamefont {Stevenson}, \citenamefont
  {Swetz}, \citenamefont {Ullom}, \citenamefont {Vale}, \citenamefont
  {Wakeham}, \citenamefont {Wassell},\ and\ \citenamefont {Wollack}}]{mwave}%
  \BibitemOpen
  \bibfield  {author} {\bibinfo {author} {\bibfnamefont {W.}~\bibnamefont
  {Yoon}}, \bibinfo {author} {\bibfnamefont {J.~S.}\ \bibnamefont {Adams}},
  \bibinfo {author} {\bibfnamefont {S.~R.}\ \bibnamefont {Bandler}}, \bibinfo
  {author} {\bibfnamefont {D.}~\bibnamefont {Becker}}, \bibinfo {author}
  {\bibfnamefont {D.~A.}\ \bibnamefont {Bennett4}}, \bibinfo {author}
  {\bibfnamefont {J.~A.}\ \bibnamefont {Chervenak}}, \bibinfo {author}
  {\bibfnamefont {A.~M.}\ \bibnamefont {Datesman}}, \bibinfo {author}
  {\bibfnamefont {M.~E.}\ \bibnamefont {Eckart}}, \bibinfo {author}
  {\bibfnamefont {F.~M.}\ \bibnamefont {Finkbeiner}}, \bibinfo {author}
  {\bibfnamefont {J.~W.}\ \bibnamefont {Fowler}}, \bibinfo {author}
  {\bibfnamefont {J.~D.}\ \bibnamefont {Gard}}, \bibinfo {author}
  {\bibfnamefont {G.~C.}\ \bibnamefont {Hilton}}, \bibinfo {author}
  {\bibfnamefont {R.~L.}\ \bibnamefont {Kelley}}, \bibinfo {author}
  {\bibfnamefont {C.~A.}\ \bibnamefont {Kilbourne}}, \bibinfo {author}
  {\bibfnamefont {J.~A.~B.}\ \bibnamefont {Mates}}, \bibinfo {author}
  {\bibfnamefont {A.~R.}\ \bibnamefont {Miniussi}}, \bibinfo {author}
  {\bibfnamefont {S.~H.}\ \bibnamefont {Moseley}}, \bibinfo {author}
  {\bibfnamefont {O.}~\bibnamefont {Noroozian}}, \bibinfo {author}
  {\bibfnamefont {F.~S.}\ \bibnamefont {Porter}}, \bibinfo {author}
  {\bibfnamefont {C.~D.}\ \bibnamefont {Reintsema}}, \bibinfo {author}
  {\bibfnamefont {J.~E.}\ \bibnamefont {Sadleir}}, \bibinfo {author}
  {\bibfnamefont {K.}~\bibnamefont {Sakai}}, \bibinfo {author} {\bibfnamefont
  {S.~J.}\ \bibnamefont {Smith}}, \bibinfo {author} {\bibfnamefont {T.~R.}\
  \bibnamefont {Stevenson}}, \bibinfo {author} {\bibfnamefont {D.~S.}\
  \bibnamefont {Swetz}}, \bibinfo {author} {\bibfnamefont {J.~N.}\ \bibnamefont
  {Ullom}}, \bibinfo {author} {\bibfnamefont {L.~R.}\ \bibnamefont {Vale}},
  \bibinfo {author} {\bibfnamefont {N.~A.}\ \bibnamefont {Wakeham}}, \bibinfo
  {author} {\bibfnamefont {E.~J.}\ \bibnamefont {Wassell}}, \ and\ \bibinfo
  {author} {\bibfnamefont {E.~J.}\ \bibnamefont {Wollack}},\ }\href@noop {}
  {\bibfield  {journal} {\bibinfo  {journal} {J. Low Temp. Phys.}\ }\textbf
  {\bibinfo {volume} {193}},\ \bibinfo {pages} {258} (\bibinfo {year}
  {2018})}\BibitemShut {NoStop}%
\bibitem [{\citenamefont {Morgan1}\ \emph {et~al.}(2016)\citenamefont
  {Morgan1}, \citenamefont {Alpert1}, \citenamefont {Bennett}, \citenamefont
  {Denison}, \citenamefont {Doriese}, \citenamefont {Fowler}, \citenamefont
  {Gard}, \citenamefont {Hilton}, \citenamefont {Irwin}, \citenamefont {Joe},
  \citenamefont {O'Neil}, \citenamefont {Reintsema}, \citenamefont {Schmidt},
  \citenamefont {Ullom},\ and\ \citenamefont {Swetz}}]{cdm}%
  \BibitemOpen
  \bibfield  {author} {\bibinfo {author} {\bibfnamefont {K.~M.}\ \bibnamefont
  {Morgan1}}, \bibinfo {author} {\bibfnamefont {B.~K.}\ \bibnamefont
  {Alpert1}}, \bibinfo {author} {\bibfnamefont {D.~A.}\ \bibnamefont
  {Bennett}}, \bibinfo {author} {\bibfnamefont {E.~V.}\ \bibnamefont
  {Denison}}, \bibinfo {author} {\bibfnamefont {W.~B.}\ \bibnamefont
  {Doriese}}, \bibinfo {author} {\bibfnamefont {J.~W.}\ \bibnamefont {Fowler}},
  \bibinfo {author} {\bibfnamefont {J.~D.}\ \bibnamefont {Gard}}, \bibinfo
  {author} {\bibfnamefont {G.~C.}\ \bibnamefont {Hilton}}, \bibinfo {author}
  {\bibfnamefont {K.~D.}\ \bibnamefont {Irwin}}, \bibinfo {author}
  {\bibfnamefont {Y.~I.}\ \bibnamefont {Joe}}, \bibinfo {author} {\bibfnamefont
  {G.~C.}\ \bibnamefont {O'Neil}}, \bibinfo {author} {\bibfnamefont {C.~D.}\
  \bibnamefont {Reintsema}}, \bibinfo {author} {\bibfnamefont {D.~R.}\
  \bibnamefont {Schmidt}}, \bibinfo {author} {\bibfnamefont {J.~N.}\
  \bibnamefont {Ullom}}, \ and\ \bibinfo {author} {\bibfnamefont {D.~S.}\
  \bibnamefont {Swetz}},\ }\href@noop {} {\bibfield  {journal} {\bibinfo
  {journal} {Appl. Phys. Lett.}\ }\textbf {\bibinfo {volume} {109}},\ \bibinfo
  {pages} {112604} (\bibinfo {year} {2016})}\BibitemShut {NoStop}%
\bibitem [{\citenamefont {Doriese}\ \emph {et~al.}(2016)\citenamefont
  {Doriese}, \citenamefont {Morgan}, \citenamefont {Bennett}, \citenamefont
  {Denison}, \citenamefont {Fitzgerald}, \citenamefont {Fowler}, \citenamefont
  {Gard}, \citenamefont {Hays-Wehle}, \citenamefont {Hilton}, \citenamefont
  {Irwin}, \citenamefont {Joe}, \citenamefont {Mates}, \citenamefont {O'Neil},
  \citenamefont {Reintsema}, \citenamefont {Robbins}, \citenamefont {Schmidt},
  \citenamefont {Swetz}, \citenamefont {Tatsuno}, \citenamefont {Vale},\ and\
  \citenamefont {Ullom}}]{tdm}%
  \BibitemOpen
  \bibfield  {author} {\bibinfo {author} {\bibfnamefont {W.~B.}\ \bibnamefont
  {Doriese}}, \bibinfo {author} {\bibfnamefont {K.~M.}\ \bibnamefont {Morgan}},
  \bibinfo {author} {\bibfnamefont {D.~A.}\ \bibnamefont {Bennett}}, \bibinfo
  {author} {\bibfnamefont {E.~V.}\ \bibnamefont {Denison}}, \bibinfo {author}
  {\bibfnamefont {C.~P.}\ \bibnamefont {Fitzgerald}}, \bibinfo {author}
  {\bibfnamefont {J.~W.}\ \bibnamefont {Fowler}}, \bibinfo {author}
  {\bibfnamefont {J.~D.}\ \bibnamefont {Gard}}, \bibinfo {author}
  {\bibfnamefont {J.~P.}\ \bibnamefont {Hays-Wehle}}, \bibinfo {author}
  {\bibfnamefont {G.~C.}\ \bibnamefont {Hilton}}, \bibinfo {author}
  {\bibfnamefont {K.~D.}\ \bibnamefont {Irwin}}, \bibinfo {author}
  {\bibfnamefont {Y.~I.}\ \bibnamefont {Joe}}, \bibinfo {author} {\bibfnamefont
  {J.~A.~B.}\ \bibnamefont {Mates}}, \bibinfo {author} {\bibfnamefont {G.~C.}\
  \bibnamefont {O'Neil}}, \bibinfo {author} {\bibfnamefont {C.~D.}\
  \bibnamefont {Reintsema}}, \bibinfo {author} {\bibfnamefont {N.~O.}\
  \bibnamefont {Robbins}}, \bibinfo {author} {\bibfnamefont {D.~R.}\
  \bibnamefont {Schmidt}}, \bibinfo {author} {\bibfnamefont {D.~S.}\
  \bibnamefont {Swetz}}, \bibinfo {author} {\bibfnamefont {H.}~\bibnamefont
  {Tatsuno}}, \bibinfo {author} {\bibfnamefont {L.~R.}\ \bibnamefont {Vale}}, \
  and\ \bibinfo {author} {\bibfnamefont {J.~N.}\ \bibnamefont {Ullom}},\
  }\href@noop {} {\bibfield  {journal} {\bibinfo  {journal} {J. Low Temp.
  Phys.}\ }\textbf {\bibinfo {volume} {184}},\ \bibinfo {pages} {389} (\bibinfo
  {year} {2016})}\BibitemShut {NoStop}%
\bibitem [{\citenamefont {Sadleir}\ \emph {et~al.}(2011)\citenamefont
  {Sadleir}, \citenamefont {Smith}, \citenamefont {Robinson}, \citenamefont
  {Finkbeiner}, \citenamefont {Chervenak}, \citenamefont {Bandler},
  \citenamefont {Eckart},\ and\ \citenamefont {Kilbourne}}]{sadleir}%
  \BibitemOpen
  \bibfield  {author} {\bibinfo {author} {\bibfnamefont {J.~E.}\ \bibnamefont
  {Sadleir}}, \bibinfo {author} {\bibfnamefont {S.~J.}\ \bibnamefont {Smith}},
  \bibinfo {author} {\bibfnamefont {I.~K.}\ \bibnamefont {Robinson}}, \bibinfo
  {author} {\bibfnamefont {F.~M.}\ \bibnamefont {Finkbeiner}}, \bibinfo
  {author} {\bibfnamefont {J.~A.}\ \bibnamefont {Chervenak}}, \bibinfo {author}
  {\bibfnamefont {S.~R.}\ \bibnamefont {Bandler}}, \bibinfo {author}
  {\bibfnamefont {M.~E.}\ \bibnamefont {Eckart}}, \ and\ \bibinfo {author}
  {\bibfnamefont {C.~A.}\ \bibnamefont {Kilbourne}},\ }\href@noop {} {\bibfield
   {journal} {\bibinfo  {journal} {Phys. Rev. B}\ }\textbf {\bibinfo {volume}
  {84}},\ \bibinfo {pages} {184502} (\bibinfo {year} {2011})}\BibitemShut
  {NoStop}%
\bibitem [{\citenamefont {Ridder}\ \emph {et~al.}(2020)\citenamefont {Ridder},
  \citenamefont {Nagayoshi}, \citenamefont {Bruijn}, \citenamefont {Gottardi},
  \citenamefont {Taralli}, \citenamefont {Khosropanah}, \citenamefont
  {Akamatsu}, \citenamefont {van~der Kuur}, \citenamefont {Ravensberg},
  \citenamefont {Visser}, \citenamefont {Nieuwenhuizen}, \citenamefont {Gao},\
  and\ \citenamefont {den Herder}}]{ridder}%
  \BibitemOpen
  \bibfield  {author} {\bibinfo {author} {\bibfnamefont {M.~L.}\ \bibnamefont
  {Ridder}}, \bibinfo {author} {\bibfnamefont {K.}~\bibnamefont {Nagayoshi}},
  \bibinfo {author} {\bibfnamefont {M.~P.}\ \bibnamefont {Bruijn}}, \bibinfo
  {author} {\bibfnamefont {L.}~\bibnamefont {Gottardi}}, \bibinfo {author}
  {\bibfnamefont {E.}~\bibnamefont {Taralli}}, \bibinfo {author} {\bibfnamefont
  {P.}~\bibnamefont {Khosropanah}}, \bibinfo {author} {\bibfnamefont
  {H.}~\bibnamefont {Akamatsu}}, \bibinfo {author} {\bibfnamefont
  {J.}~\bibnamefont {van~der Kuur}}, \bibinfo {author} {\bibfnamefont
  {K.}~\bibnamefont {Ravensberg}}, \bibinfo {author} {\bibfnamefont
  {S.}~\bibnamefont {Visser}}, \bibinfo {author} {\bibfnamefont {A.~C.~T.}\
  \bibnamefont {Nieuwenhuizen}}, \bibinfo {author} {\bibfnamefont {J.~R.}\
  \bibnamefont {Gao}}, \ and\ \bibinfo {author} {\bibfnamefont {J.-W.}\
  \bibnamefont {den Herder}},\ }\href@noop {} {\bibfield  {journal} {\bibinfo
  {journal} {J. Low Temp. Phys.}\ }\textbf {\bibinfo {volume} {199}},\ \bibinfo
  {pages} {962} (\bibinfo {year} {2020})}\BibitemShut {NoStop}%
\bibitem [{\citenamefont {Smith}\ \emph {et~al.}(2013)\citenamefont {Smith},
  \citenamefont {Adams}, \citenamefont {Bailey}, \citenamefont {Bandler},
  \citenamefont {Busch}, \citenamefont {Chervenak}, \citenamefont {Eckart},
  \citenamefont {Finkbeiner}, \citenamefont {Kilbourne}, \citenamefont
  {Kelley}, \citenamefont {Lee}, \citenamefont {Porst}, \citenamefont
  {Porter},\ and\ \citenamefont {Sadleir}}]{smith}%
  \BibitemOpen
  \bibfield  {author} {\bibinfo {author} {\bibfnamefont {S.~J.}\ \bibnamefont
  {Smith}}, \bibinfo {author} {\bibfnamefont {J.~S.}\ \bibnamefont {Adams}},
  \bibinfo {author} {\bibfnamefont {C.~N.}\ \bibnamefont {Bailey}}, \bibinfo
  {author} {\bibfnamefont {S.~R.}\ \bibnamefont {Bandler}}, \bibinfo {author}
  {\bibfnamefont {S.~E.}\ \bibnamefont {Busch}}, \bibinfo {author}
  {\bibfnamefont {J.~A.}\ \bibnamefont {Chervenak}}, \bibinfo {author}
  {\bibfnamefont {M.~E.}\ \bibnamefont {Eckart}}, \bibinfo {author}
  {\bibfnamefont {F.~M.}\ \bibnamefont {Finkbeiner}}, \bibinfo {author}
  {\bibfnamefont {C.~A.}\ \bibnamefont {Kilbourne}}, \bibinfo {author}
  {\bibfnamefont {R.~L.}\ \bibnamefont {Kelley}}, \bibinfo {author}
  {\bibfnamefont {S.-J.}\ \bibnamefont {Lee}}, \bibinfo {author} {\bibfnamefont
  {J.-P.}\ \bibnamefont {Porst}}, \bibinfo {author} {\bibfnamefont {F.~S.}\
  \bibnamefont {Porter}}, \ and\ \bibinfo {author} {\bibfnamefont {J.~E.}\
  \bibnamefont {Sadleir}},\ }\href@noop {} {\bibfield  {journal} {\bibinfo
  {journal} {J. Appl. Phys.}\ }\textbf {\bibinfo {volume} {114}},\ \bibinfo
  {pages} {074513} (\bibinfo {year} {2013})}\BibitemShut {NoStop}%
\bibitem [{\citenamefont {Taralli}\ \emph {et~al.}(2019)\citenamefont
  {Taralli}, \citenamefont {Khosropanah}, \citenamefont {Gottardi},
  \citenamefont {Nagayoshi}, \citenamefont {Ridder}, \citenamefont {Bruijn},\
  and\ \citenamefont {Gao}}]{taralli}%
  \BibitemOpen
  \bibfield  {author} {\bibinfo {author} {\bibfnamefont {E.}~\bibnamefont
  {Taralli}}, \bibinfo {author} {\bibfnamefont {P.}~\bibnamefont
  {Khosropanah}}, \bibinfo {author} {\bibfnamefont {L.}~\bibnamefont
  {Gottardi}}, \bibinfo {author} {\bibfnamefont {K.}~\bibnamefont {Nagayoshi}},
  \bibinfo {author} {\bibfnamefont {M.~L.}\ \bibnamefont {Ridder}}, \bibinfo
  {author} {\bibfnamefont {M.~P.}\ \bibnamefont {Bruijn}}, \ and\ \bibinfo
  {author} {\bibfnamefont {J.~R.}\ \bibnamefont {Gao}},\ }\href@noop {}
  {\bibfield  {journal} {\bibinfo  {journal} {AIP Adv.}\ }\textbf {\bibinfo
  {volume} {9}},\ \bibinfo {pages} {045324} (\bibinfo {year}
  {2019})}\BibitemShut {NoStop}%
\bibitem [{\citenamefont {Wakeham}\ \emph {et~al.}(2018)\citenamefont
  {Wakeham}, \citenamefont {Adams}, \citenamefont {Bandler}, \citenamefont
  {Chervenak}, \citenamefont {Datesman}, \citenamefont {Eckart}, \citenamefont
  {Finkbeiner}, \citenamefont {Kelley}, \citenamefont {Kilbourne},
  \citenamefont {Miniussi}, \citenamefont {Porter}, \citenamefont {Sadleir},
  \citenamefont {Sakai}, \citenamefont {Smith}, , \citenamefont {Wassell},\
  and\ \citenamefont {Yoon}}]{nick2}%
  \BibitemOpen
  \bibfield  {author} {\bibinfo {author} {\bibfnamefont {N.~A.}\ \bibnamefont
  {Wakeham}}, \bibinfo {author} {\bibfnamefont {J.~S.}\ \bibnamefont {Adams}},
  \bibinfo {author} {\bibfnamefont {S.~R.}\ \bibnamefont {Bandler}}, \bibinfo
  {author} {\bibfnamefont {J.~A.}\ \bibnamefont {Chervenak}}, \bibinfo {author}
  {\bibfnamefont {A.~M.}\ \bibnamefont {Datesman}}, \bibinfo {author}
  {\bibfnamefont {M.~E.}\ \bibnamefont {Eckart}}, \bibinfo {author}
  {\bibfnamefont {F.~M.}\ \bibnamefont {Finkbeiner}}, \bibinfo {author}
  {\bibfnamefont {R.~L.}\ \bibnamefont {Kelley}}, \bibinfo {author}
  {\bibfnamefont {C.~A.}\ \bibnamefont {Kilbourne}}, \bibinfo {author}
  {\bibfnamefont {A.~R.}\ \bibnamefont {Miniussi}}, \bibinfo {author}
  {\bibfnamefont {F.~S.}\ \bibnamefont {Porter}}, \bibinfo {author}
  {\bibfnamefont {J.~E.}\ \bibnamefont {Sadleir}}, \bibinfo {author}
  {\bibfnamefont {K.}~\bibnamefont {Sakai}}, \bibinfo {author} {\bibfnamefont
  {S.~J.}\ \bibnamefont {Smith}}, , \bibinfo {author} {\bibfnamefont {E.~J.}\
  \bibnamefont {Wassell}}, \ and\ \bibinfo {author} {\bibfnamefont
  {W.}~\bibnamefont {Yoon}},\ }\href@noop {} {\bibfield  {journal} {\bibinfo
  {journal} {J. Low. Temp. Phys.}\ }\textbf {\bibinfo {volume} {193}},\
  \bibinfo {pages} {164503} (\bibinfo {year} {2018})}\BibitemShut {NoStop}%
\bibitem [{\citenamefont {Gottardi}\ \emph {et~al.}(2018)\citenamefont
  {Gottardi}, \citenamefont {Smith}, \citenamefont {Kozorezov}, \citenamefont
  {Akamatsu}, \citenamefont {van~der Kuur}, \citenamefont {Bandler},
  \citenamefont {Bruijn}, \citenamefont {Chervenak}, \citenamefont {Gao},
  \citenamefont {den Hartog}, \citenamefont {Jackson}, \citenamefont
  {Khosropanah}, \citenamefont {Miniussi}, \citenamefont {Nagayoshi},
  \citenamefont {Ridder}, \citenamefont {Sadleir}, \citenamefont {Sakai},\ and\
  \citenamefont {Wakeham}}]{gottardi1}%
  \BibitemOpen
  \bibfield  {author} {\bibinfo {author} {\bibfnamefont {L.}~\bibnamefont
  {Gottardi}}, \bibinfo {author} {\bibfnamefont {S.~J.}\ \bibnamefont {Smith}},
  \bibinfo {author} {\bibfnamefont {A.}~\bibnamefont {Kozorezov}}, \bibinfo
  {author} {\bibfnamefont {H.}~\bibnamefont {Akamatsu}}, \bibinfo {author}
  {\bibfnamefont {J.}~\bibnamefont {van~der Kuur}}, \bibinfo {author}
  {\bibfnamefont {S.~R.}\ \bibnamefont {Bandler}}, \bibinfo {author}
  {\bibfnamefont {M.~P.}\ \bibnamefont {Bruijn}}, \bibinfo {author}
  {\bibfnamefont {J.~A.}\ \bibnamefont {Chervenak}}, \bibinfo {author}
  {\bibfnamefont {J.~R.}\ \bibnamefont {Gao}}, \bibinfo {author} {\bibfnamefont
  {R.~H.}\ \bibnamefont {den Hartog}}, \bibinfo {author} {\bibfnamefont
  {B.~D.}\ \bibnamefont {Jackson}}, \bibinfo {author} {\bibfnamefont
  {P.}~\bibnamefont {Khosropanah}}, \bibinfo {author} {\bibfnamefont
  {A.}~\bibnamefont {Miniussi}}, \bibinfo {author} {\bibfnamefont
  {K.}~\bibnamefont {Nagayoshi}}, \bibinfo {author} {\bibfnamefont
  {M.}~\bibnamefont {Ridder}}, \bibinfo {author} {\bibfnamefont
  {J.}~\bibnamefont {Sadleir}}, \bibinfo {author} {\bibfnamefont
  {K.}~\bibnamefont {Sakai}}, \ and\ \bibinfo {author} {\bibfnamefont
  {N.}~\bibnamefont {Wakeham}},\ }\href@noop {} {\bibfield  {journal} {\bibinfo
   {journal} {J. Low Temp. Phys.}\ }\textbf {\bibinfo {volume} {231?240}},\
  \bibinfo {pages} {209} (\bibinfo {year} {2018})}\BibitemShut {NoStop}%
\bibitem [{\citenamefont {Gottardi}\ \emph {et~al.}(2017)\citenamefont
  {Gottardi}, \citenamefont {Akamatsu}, \citenamefont {van~der Kuur},
  \citenamefont {Smith}, \citenamefont {Kozorezov},\ and\ \citenamefont
  {Chervenak}}]{gottardi2}%
  \BibitemOpen
  \bibfield  {author} {\bibinfo {author} {\bibfnamefont {L.}~\bibnamefont
  {Gottardi}}, \bibinfo {author} {\bibfnamefont {H.}~\bibnamefont {Akamatsu}},
  \bibinfo {author} {\bibfnamefont {J.}~\bibnamefont {van~der Kuur}}, \bibinfo
  {author} {\bibfnamefont {S.~J.}\ \bibnamefont {Smith}}, \bibinfo {author}
  {\bibfnamefont {A.}~\bibnamefont {Kozorezov}}, \ and\ \bibinfo {author}
  {\bibfnamefont {J.}~\bibnamefont {Chervenak}},\ }\href@noop {} {\bibfield
  {journal} {\bibinfo  {journal} {IEEE Transactions on Applied
  Superconductivity}\ }\textbf {\bibinfo {volume} {27}},\ \bibinfo {pages}
  {2101404} (\bibinfo {year} {2017})}\BibitemShut {NoStop}%
\bibitem [{\citenamefont {Gottardi}\ \emph {et~al.}(2014)\citenamefont
  {Gottardi}, \citenamefont {Kozorezov}, \citenamefont {Akamatsu},
  \citenamefont {van~der Kuur}, \citenamefont {Bruijn}, \citenamefont {den
  Hartog}, \citenamefont {Hijmering}, \citenamefont {Khosropanah},
  \citenamefont {Lambert}, \citenamefont {van~der Linden}, \citenamefont
  {Ridder}, \citenamefont {Suzuki}, ,\ and\ \citenamefont {Gao}}]{gottardi3}%
  \BibitemOpen
  \bibfield  {author} {\bibinfo {author} {\bibfnamefont {L.}~\bibnamefont
  {Gottardi}}, \bibinfo {author} {\bibfnamefont {A.}~\bibnamefont {Kozorezov}},
  \bibinfo {author} {\bibfnamefont {H.}~\bibnamefont {Akamatsu}}, \bibinfo
  {author} {\bibfnamefont {J.}~\bibnamefont {van~der Kuur}}, \bibinfo {author}
  {\bibfnamefont {M.~P.}\ \bibnamefont {Bruijn}}, \bibinfo {author}
  {\bibfnamefont {R.~H.}\ \bibnamefont {den Hartog}}, \bibinfo {author}
  {\bibfnamefont {R.}~\bibnamefont {Hijmering}}, \bibinfo {author}
  {\bibfnamefont {P.}~\bibnamefont {Khosropanah}}, \bibinfo {author}
  {\bibfnamefont {C.}~\bibnamefont {Lambert}}, \bibinfo {author} {\bibfnamefont
  {A.~J.}\ \bibnamefont {van~der Linden}}, \bibinfo {author} {\bibfnamefont
  {M.~L.}\ \bibnamefont {Ridder}}, \bibinfo {author} {\bibfnamefont
  {T.}~\bibnamefont {Suzuki}}, , \ and\ \bibinfo {author} {\bibfnamefont
  {J.~R.}\ \bibnamefont {Gao}},\ }\href@noop {} {\bibfield  {journal} {\bibinfo
   {journal} {Appl. Phys. Lett.}\ }\textbf {\bibinfo {volume} {1105}},\
  \bibinfo {pages} {162605} (\bibinfo {year} {2014})}\BibitemShut {NoStop}%
\bibitem [{\citenamefont {Wakeham}\ \emph {et~al.}(2019)\citenamefont
  {Wakeham}, \citenamefont {Adams}, \citenamefont {Beaumont}, \citenamefont
  {Chervenak}, \citenamefont {Datesman}, \citenamefont {Eckart}, \citenamefont
  {Finkbeiner}, \citenamefont {Hummatov}, \citenamefont {Kelley}, \citenamefont
  {Kilbourne}, \citenamefont {Miniussi}, \citenamefont {Porter}, \citenamefont
  {Sadleir}, \citenamefont {Sakai}, \citenamefont {Smith}, ,\ and\
  \citenamefont {Wassell}}]{nick}%
  \BibitemOpen
  \bibfield  {author} {\bibinfo {author} {\bibfnamefont {N.~A.}\ \bibnamefont
  {Wakeham}}, \bibinfo {author} {\bibfnamefont {J.~S.}\ \bibnamefont {Adams}},
  \bibinfo {author} {\bibfnamefont {S.~R. B.~S.}\ \bibnamefont {Beaumont}},
  \bibinfo {author} {\bibfnamefont {J.~A.}\ \bibnamefont {Chervenak}}, \bibinfo
  {author} {\bibfnamefont {A.~M.}\ \bibnamefont {Datesman}}, \bibinfo {author}
  {\bibfnamefont {M.~E.}\ \bibnamefont {Eckart}}, \bibinfo {author}
  {\bibfnamefont {F.~M.}\ \bibnamefont {Finkbeiner}}, \bibinfo {author}
  {\bibfnamefont {R.}~\bibnamefont {Hummatov}}, \bibinfo {author}
  {\bibfnamefont {R.~L.}\ \bibnamefont {Kelley}}, \bibinfo {author}
  {\bibfnamefont {C.~A.}\ \bibnamefont {Kilbourne}}, \bibinfo {author}
  {\bibfnamefont {A.~R.}\ \bibnamefont {Miniussi}}, \bibinfo {author}
  {\bibfnamefont {F.~S.}\ \bibnamefont {Porter}}, \bibinfo {author}
  {\bibfnamefont {J.~E.}\ \bibnamefont {Sadleir}}, \bibinfo {author}
  {\bibfnamefont {K.}~\bibnamefont {Sakai}}, \bibinfo {author} {\bibfnamefont
  {S.~J.}\ \bibnamefont {Smith}}, , \ and\ \bibinfo {author} {\bibfnamefont
  {E.~J.}\ \bibnamefont {Wassell}},\ }\href@noop {} {\bibfield  {journal}
  {\bibinfo  {journal} {J. Appl. Phys.}\ }\textbf {\bibinfo {volume} {125}},\
  \bibinfo {pages} {164503} (\bibinfo {year} {2019})}\BibitemShut {NoStop}%
\bibitem [{\citenamefont {Nagayoshi}\ \emph {et~al.}(2020)\citenamefont
  {Nagayoshi}, \citenamefont {Ridder}, \citenamefont {Bruijn}, \citenamefont
  {Gottardi}, \citenamefont {Taralli}, \citenamefont {Khosropanah},
  \citenamefont {Akamatsu1}, \citenamefont {Visser},\ and\ \citenamefont
  {Gao}}]{ken}%
  \BibitemOpen
  \bibfield  {author} {\bibinfo {author} {\bibfnamefont {K.}~\bibnamefont
  {Nagayoshi}}, \bibinfo {author} {\bibfnamefont {M.~L.}\ \bibnamefont
  {Ridder}}, \bibinfo {author} {\bibfnamefont {M.~P.}\ \bibnamefont {Bruijn}},
  \bibinfo {author} {\bibfnamefont {L.}~\bibnamefont {Gottardi}}, \bibinfo
  {author} {\bibfnamefont {E.}~\bibnamefont {Taralli}}, \bibinfo {author}
  {\bibfnamefont {P.}~\bibnamefont {Khosropanah}}, \bibinfo {author}
  {\bibfnamefont {H.}~\bibnamefont {Akamatsu1}}, \bibinfo {author}
  {\bibfnamefont {S.}~\bibnamefont {Visser}}, \ and\ \bibinfo {author}
  {\bibfnamefont {J.~R.}\ \bibnamefont {Gao}},\ }\href@noop {} {\bibfield
  {journal} {\bibinfo  {journal} {J. Low Temp. Phys.}\ }\textbf {\bibinfo
  {volume} {199}},\ \bibinfo {pages} {943} (\bibinfo {year}
  {2020})}\BibitemShut {NoStop}%
\bibitem [{\citenamefont {Taralli}\ \emph {et~al.}(2020)\citenamefont
  {Taralli}, \citenamefont {Gottardi}, \citenamefont {Nagayoshi}, \citenamefont
  {Ridder}, \citenamefont {Visser}, \citenamefont {Khosropanah}, \citenamefont
  {Akamatsu}, \citenamefont {van~der Kuur}, \citenamefont {Bruijn},\ and\
  \citenamefont {Gao}}]{taralli2}%
  \BibitemOpen
  \bibfield  {author} {\bibinfo {author} {\bibfnamefont {E.}~\bibnamefont
  {Taralli}}, \bibinfo {author} {\bibfnamefont {L.}~\bibnamefont {Gottardi}},
  \bibinfo {author} {\bibfnamefont {K.}~\bibnamefont {Nagayoshi}}, \bibinfo
  {author} {\bibfnamefont {M.}~\bibnamefont {Ridder}}, \bibinfo {author}
  {\bibfnamefont {S.}~\bibnamefont {Visser}}, \bibinfo {author} {\bibfnamefont
  {P.}~\bibnamefont {Khosropanah}}, \bibinfo {author} {\bibfnamefont
  {H.}~\bibnamefont {Akamatsu}}, \bibinfo {author} {\bibfnamefont
  {J.}~\bibnamefont {van~der Kuur}}, \bibinfo {author} {\bibfnamefont
  {M.}~\bibnamefont {Bruijn}}, \ and\ \bibinfo {author} {\bibfnamefont {J.~R.}\
  \bibnamefont {Gao}},\ }\href@noop {} {\bibfield  {journal} {\bibinfo
  {journal} {J. Low Temp. Phys.}\ }\textbf {\bibinfo {volume} {199}},\ \bibinfo
  {pages} {80} (\bibinfo {year} {2020})}\BibitemShut {NoStop}%
\bibitem [{\citenamefont {de~Wit}\ \emph {et~al.}(2020)\citenamefont {de~Wit},
  \citenamefont {Gottardi}, \citenamefont {Taralli}, \citenamefont {Nagayoshi},
  \citenamefont {Ridder}, \citenamefont {Akamatsu}, \citenamefont {Ravensberg},
  \citenamefont {D'Andrea}, \citenamefont {Vaccaro}, \citenamefont {Visser},
  \citenamefont {Bruijn},\ and\ \citenamefont {Gao}}]{martin}%
  \BibitemOpen
  \bibfield  {author} {\bibinfo {author} {\bibfnamefont {M.}~\bibnamefont
  {de~Wit}}, \bibinfo {author} {\bibfnamefont {L.}~\bibnamefont {Gottardi}},
  \bibinfo {author} {\bibfnamefont {E.}~\bibnamefont {Taralli}}, \bibinfo
  {author} {\bibfnamefont {K.}~\bibnamefont {Nagayoshi}}, \bibinfo {author}
  {\bibfnamefont {M.}~\bibnamefont {Ridder}}, \bibinfo {author} {\bibfnamefont
  {H.}~\bibnamefont {Akamatsu}}, \bibinfo {author} {\bibfnamefont
  {K.}~\bibnamefont {Ravensberg}}, \bibinfo {author} {\bibfnamefont
  {M.}~\bibnamefont {D'Andrea}}, \bibinfo {author} {\bibfnamefont
  {D.}~\bibnamefont {Vaccaro}}, \bibinfo {author} {\bibfnamefont
  {S.}~\bibnamefont {Visser}}, \bibinfo {author} {\bibfnamefont
  {M.}~\bibnamefont {Bruijn}}, \ and\ \bibinfo {author} {\bibfnamefont
  {J.}~\bibnamefont {Gao}},\ }\href@noop {} {\bibfield  {journal} {\bibinfo
  {journal} {J. Appl. Phys.}\ }\textbf {\bibinfo {volume} {128}},\ \bibinfo
  {pages} {doi: 10.1063/5.0029669} (\bibinfo {year} {2020})}\BibitemShut
  {NoStop}%
\bibitem [{\citenamefont {Hoevers}\ \emph {et~al.}(2005)\citenamefont
  {Hoevers}, \citenamefont {Ridder}, \citenamefont {Germeaui}, \citenamefont
  {Bruijn}, \citenamefont {de~Korte},\ and\ \citenamefont
  {Wiegerink}}]{thermalG1}%
  \BibitemOpen
  \bibfield  {author} {\bibinfo {author} {\bibfnamefont {H.~F.~C.}\
  \bibnamefont {Hoevers}}, \bibinfo {author} {\bibfnamefont {M.~L.}\
  \bibnamefont {Ridder}}, \bibinfo {author} {\bibfnamefont {A.}~\bibnamefont
  {Germeaui}}, \bibinfo {author} {\bibfnamefont {M.~P.}\ \bibnamefont
  {Bruijn}}, \bibinfo {author} {\bibfnamefont {P.~A.~J.}\ \bibnamefont
  {de~Korte}}, \ and\ \bibinfo {author} {\bibfnamefont {R.~J.}\ \bibnamefont
  {Wiegerink}},\ }\href@noop {} {\bibfield  {journal} {\bibinfo  {journal}
  {Appl. Phys. Lett.}\ }\textbf {\bibinfo {volume} {86}},\ \bibinfo {pages}
  {251903,} (\bibinfo {year} {2005})}\BibitemShut {NoStop}%
\bibitem [{foo()}]{footnote}%
  \BibitemOpen
  \href@noop {} {}\bibinfo {note} {We use our electronics also to read out the
  resistance of our thermometer. Due to a different calibration among warm
  electronics, we read different resistance value out of our thermometer and
  thus we measured different $T_\mathrm{c}$ in Run 1. The other measurements
  are not affected by this problem: energy resolution and NEP are calibrated by
  the X-ray pulse energy while the complex impedance measurement uses a
  self-calibrating method taking into account the transfer function of the
  entire system}\BibitemShut {NoStop}%
\bibitem [{\citenamefont {Kilbourne}\ \emph {et~al.}(2007)\citenamefont
  {Kilbourne}, \citenamefont {Bandler}, \citenamefont {Brown}, \citenamefont
  {Chervenak}, \citenamefont {Figueroa-Feliciano}, \citenamefont {Finkbeiner},
  \citenamefont {Iyomoto}, \citenamefont {Kelley}, \citenamefont {Porter},\
  and\ \citenamefont {Smith}}]{thermalG2}%
  \BibitemOpen
  \bibinfo {editor} {\bibfnamefont {C.~A.}\ \bibnamefont {Kilbourne}}, \bibinfo
  {editor} {\bibfnamefont {S.~R.}\ \bibnamefont {Bandler}}, \bibinfo {editor}
  {\bibfnamefont {A.~D.}\ \bibnamefont {Brown}}, \bibinfo {editor}
  {\bibfnamefont {J.~A.}\ \bibnamefont {Chervenak}}, \bibinfo {editor}
  {\bibfnamefont {E.}~\bibnamefont {Figueroa-Feliciano}}, \bibinfo {editor}
  {\bibfnamefont {F.~M.}\ \bibnamefont {Finkbeiner}}, \bibinfo {editor}
  {\bibfnamefont {N.}~\bibnamefont {Iyomoto}}, \bibinfo {editor} {\bibfnamefont
  {R.~L.}\ \bibnamefont {Kelley}}, \bibinfo {editor} {\bibfnamefont {F.~S.}\
  \bibnamefont {Porter}}, \ and\ \bibinfo {editor} {\bibfnamefont {S.~J.}\
  \bibnamefont {Smith}},\ eds.,\ \href@noop {} {\emph {\bibinfo {title} {Proc.
  SPIE}}},\ Vol.\ \bibinfo {volume} {6686}\ (\bibinfo {year}
  {2007})\BibitemShut {NoStop}%
\bibitem [{\citenamefont {Lindeman}\ \emph {et~al.}(2004)\citenamefont
  {Lindeman}, \citenamefont {Bandler}, \citenamefont {Brekosky}, \citenamefont
  {Chervenak}, \citenamefont {F.igueroa-Feliciano}, \citenamefont {Finkbeiner},
  \citenamefont {Li},\ and\ \citenamefont {Kilbourne}}]{lindeman}%
  \BibitemOpen
  \bibfield  {author} {\bibinfo {author} {\bibfnamefont {M.~A.}\ \bibnamefont
  {Lindeman}}, \bibinfo {author} {\bibfnamefont {S.}~\bibnamefont {Bandler}},
  \bibinfo {author} {\bibfnamefont {R.~P.}\ \bibnamefont {Brekosky}}, \bibinfo
  {author} {\bibfnamefont {J.~A.}\ \bibnamefont {Chervenak}}, \bibinfo {author}
  {\bibfnamefont {E.}~\bibnamefont {F.igueroa-Feliciano}}, \bibinfo {author}
  {\bibfnamefont {F.~M.}\ \bibnamefont {Finkbeiner}}, \bibinfo {author}
  {\bibfnamefont {M.~J.}\ \bibnamefont {Li}}, \ and\ \bibinfo {author}
  {\bibfnamefont {C.~A.}\ \bibnamefont {Kilbourne}},\ }\href@noop {} {\bibfield
   {journal} {\bibinfo  {journal} {Rev. Sci. Instrum.}\ }\textbf {\bibinfo
  {volume} {75}},\ \bibinfo {pages} {1283} (\bibinfo {year}
  {2004})}\BibitemShut {NoStop}%
\bibitem [{\citenamefont {Taralli.}\ \emph {et~al.}(2010)\citenamefont
  {Taralli.}, \citenamefont {Portesi}, \citenamefont {Lolli}, \citenamefont
  {Rajteri}, \citenamefont {Monticone}, \citenamefont {Novikov},\ and\
  \citenamefont {Beyer}}]{taralli3}%
  \BibitemOpen
  \bibfield  {author} {\bibinfo {author} {\bibfnamefont {E.}~\bibnamefont
  {Taralli.}}, \bibinfo {author} {\bibfnamefont {C.}~\bibnamefont {Portesi}},
  \bibinfo {author} {\bibfnamefont {L.}~\bibnamefont {Lolli}}, \bibinfo
  {author} {\bibfnamefont {M.}~\bibnamefont {Rajteri}}, \bibinfo {author}
  {\bibfnamefont {E.}~\bibnamefont {Monticone}}, \bibinfo {author}
  {\bibfnamefont {I.}~\bibnamefont {Novikov}}, \ and\ \bibinfo {author}
  {\bibfnamefont {J.}~\bibnamefont {Beyer}},\ }\href@noop {} {\bibfield
  {journal} {\bibinfo  {journal} {Supercond. Sci. Technol.}\ }\textbf {\bibinfo
  {volume} {23}},\ \bibinfo {pages} {105012} (\bibinfo {year}
  {2010})}\BibitemShut {NoStop}%
\bibitem [{\citenamefont {Hattori1}\ \emph {et~al.}(2020)\citenamefont
  {Hattori1}, \citenamefont {Kobayashi}, \citenamefont {Takasu},\ and\
  \citenamefont {Fukuda}}]{fukuda2}%
  \BibitemOpen
  \bibfield  {author} {\bibinfo {author} {\bibfnamefont {K.}~\bibnamefont
  {Hattori1}}, \bibinfo {author} {\bibfnamefont {R.}~\bibnamefont {Kobayashi}},
  \bibinfo {author} {\bibfnamefont {S.}~\bibnamefont {Takasu}}, \ and\ \bibinfo
  {author} {\bibfnamefont {D.}~\bibnamefont {Fukuda}},\ }\href@noop {}
  {\bibfield  {journal} {\bibinfo  {journal} {AIP Adv.}\ }\textbf {\bibinfo
  {volume} {10}},\ \bibinfo {pages} {035004} (\bibinfo {year}
  {2020})}\BibitemShut {NoStop}%
\end{thebibliography}%
 
\end{document}